\documentclass[3p, times]{elsarticle}

\usepackage{rotating}
\usepackage{amssymb}                       
\usepackage{amsmath}                       
\usepackage{amsthm}                        
\usepackage{listings}                      
\usepackage{subcaption}
\usepackage{graphicx}
\usepackage{color,soul}
\usepackage{algpseudocode}
\usepackage{multirow}
\usepackage{algorithm}
\usepackage[utf8]{inputenc}
\usepackage{hyperref}

\journal{Computer Physics Communications}

\newcommand{\be}{\begin{equation}}
\newcommand{\ee}{\end{equation}}
\newcommand{\bi}{\begin{itemize}}
\newcommand{\ei}{\end{itemize}}

\newcommand{\dd}[2]{{\frac{\partial #1}{\partial #2}}}

\newcommand{\beq}{\begin{equation}}
\newcommand{\eeq}{\end{equation}}

\begin{document}

\begin{frontmatter}

\title{A GPU-enabled implicit Finite Volume solver for the ideal two-fluid plasma model on unstructured grids}

\author[lagunaaddress]{Isaac Alonso Asensio\corref{mycorrespondingauthor}}
\cortext[mycorrespondingauthor]{Corresponding author}
\ead{isaac.alonso.07@ull.edu.es}

\author[lppaddress]{Alejandro Alvarez Laguna}

\author[vkiaddress]{Mohamed Hassanine Aissa}

\author[kuladdress]{Stefaan Poedts}

\author[kuladdress]{Nataly Ozak}

\author[kuladdress]{Andrea Lani}

\address[lagunaaddress]{University of La Laguna, Avda. Astrofísico Francisco Sánchez 38206, La Laguna, Tenerife, Spain}
\address[vkiaddress]{Von Karman Institute for Fluid Dynamics, Waterloosesteenweg 72, 1640, Sint Genesius Rode, Belgium}
\address[lppaddress]{Laboratoire de Physique des plasmas, Ecole Polytechnique,91128 PALAISEAU CEDEX, France}
\address[kuladdress]{KU Leuven/Centrum voor mathematische Plasma-Astrofysica, Celestijnenlaan 200B, B-3001 Leuven, Belgium} 

\begin{abstract}

This paper describes the main features of a pioneering unsteady solver for simulating ideal two-fluid plasmas on unstructured grids, taking profit of GPGPU (General-purpose computing on graphics processing units). The code, which has been implemented within the open source COOLFluiD platform, is implicit, second-order in time and space, relying upon a Finite Volume method for the spatial discretization and a three-point backward Euler for the time integration. In particular, the convective fluxes are computed by a multi-fluid version of the AUSM+up scheme for the plasma equations, in combination with a modified Rusanov scheme with tunable dissipation for the Maxwell equations. Source terms are integrated with a one-point rule, using the cell-centered value. Some critical aspects of the porting to GPU's are discussed, as well as the performance of two open source linear system solvers (i.e. PETSc, PARALUTION). The code design allows for computing both flux and source terms on the GPU along with their Jacobian, giving a noticeable decrease in the computational time in comparison with the original CPU-based solver. The code has been tested in a wide range of mesh sizes and in three different systems, each one with a different GPU. The increased performance (up to 14x) is demonstrated in two representative 2D benchmarks: propagation of circularly polarized waves and the more challenging Geospace Environmental Modeling (GEM) magnetic reconnection challenge.

\end{abstract}

\begin{keyword}
Magnetohydrodynamics \sep Finite Volume \sep GPU \sep object-oriented \sep multi-fluids \sep plasmas
\end{keyword}

\end{frontmatter}


\section*{Introduction}

The simulations of heliospheric phenomena, in particular those driving Space Weather which are of motivating interest for this work, need, ideally, to be performed faster-than-real-time in order to be useful for forecasting purposes. Those kind of numerical predictions currently rely, at best, upon a single-fluid Magnetohydrodynamic (MHD) approach. The latter, however, presents some clear limitations on the physics that can be simulated, as it cannot take into account rigorously the multi-component nature of the solar/magnetospheric plasma. Therefore, most of the MHD models neglect weak ionization effects (where applicable), including collisional effects, chemical reactions, presence of neutrals, the different dynamics between ions and electrons or non-LTE radiation. In order to potentially tackle all those effects and better characterize phenomena which are induced by space plasmas, novel multi-fluid models have been recently developed \cite{Shumlak11, AlvarezLaguna16, maneva17,AlvarezLaguna18-CPC,AlvarezLaguna18, Amano15}. In those models, each plasma component (electrons, ions, neutrals) is treated independently as a separate fluid including a coupling to the electromagnetic field through the Lorentz force and the Joule heating term, while the interactions between different fluids are modeled as chemical reactions or collisions. \\

When simulating those augmented systems of partial differential equations (PDE's), much more complex and computationally demanding than standard MHD, it becomes mandatory to seek ways to boost run-time performance as much as possible, exploiting at best modern High Performance Computing (HPC) systems, such as those featuring both multiple CPU-cores and accelerators like Graphical Processing Units (GPU) \cite{Owens08}. Since a few years, a slow adaptation of scientific codes to the so-called heterogeneous systems have started, but, unlike for other scientific fields (e.g. N-body simulations \cite{Potter17}, chemistry \cite{Ufimtsev08}, molecular dynamics \cite{Anderson08}), this effort has been progressing rather slowly and, often, without fully satisfactory performance gain for implicit plasma codes. Among some possible reasons for this relatively slow process, at least up to now, we can cite two: (1) the complexity associated to porting flow codes to GPU architectures, which in some cases may require a substantial redesign of computational kernels in order to get some significant return in performance, and (2) the limited performance benefit of linear system solvers and, in particular, parallel preconditioners (both needed by time-implicit codes) when ported to GPU's \cite{Li2013}. \\

Unlike CPUs, accelerators such as GPUs have limited local memory of the order of a few GBs for a single GPU (e.g. the NVIDIA Tesla K40 has 12 GB). Even though modern GPUs are progressing slightly on this specification, it remains few orders of magnitudes below what CPUs offers in terms of memory. Because of a low memory bandwidth, the communication between the CPU and the GPU has to be kept minimal and the GPU should be independent of the host CPU for most of the simulation.  Therefore, memory limitation is present while designing new algorithms for the GPU.
For CFD simulations, the memory limitation is directly translated into a maximum size of the mesh that can be supported on a single GPU device \cite{Aissa2017}. In order to handle larger meshes implicit solvers could make use of multiple GPUs \cite{Watkins16}. \\

Memory-greedy algorithms such as linear system solvers are challenging for the GPU. While an explicit time stepping requires the storage of the mesh, the residual and the solution, an implicit time stepping stores additionally a large sparse system matrix (including the Jacobian of the full discretization), the preconditioner (which can be a sparse matrix as well, e.g. in ILU methods) and the time step.
This system matrix, which depends on the number of cells and the number of variables of the plasma solver, easily takes a few GBs for a fine 2D unstructured mesh, as in the GEM case.
A GPU-friendly way to implement implicit solver would be to use a matrix-free algorithm for which however a considerable code refactoring is needed.
Another alternative is to use slicing and solve the problem slice per slice which is adapted especially for multiblock meshes, but this is not trivial to implement for unstructured meshes as in the present case.
Some linear system solvers store many intermediate results which are used to converge to the final solution. Such an approach is not suited for the GPU since the memory consumption is proportional to the number of linear iterations which could exceed a few hundred for stiff problems. In the current work, the Generalized Minimal RESidual (GMRES) algorithms are used to solve the linear systems but with a restart feature to limit the memory utilization. The linear solver stores indeed only a limited number of intermediate results, which represent a compromise between linear solver convergence and memory limitation. Moreover, the Jacobi preconditioner, used in this work, is less memory greedy than others such as ILU and therefore more suited for the GPU context.
Even for large clusters of powerful CPUs, memory consumption constitutes a strong selection criterion, especially for large-scale problems. Lowering memory consumption is thus a good practice to design new algorithms for HPC architecture. \\

The present paper addresses the previously described aspects by demonstrating, respectively, that (1) an object-oriented approach can substantially ease the porting of complex physics codes (in this case, multi-fluid plasma, unstructured, implicit time-accurate), and (2) that recently developed open source GPU-enabled linear system solvers are getting much more efficient and competitive than their solely CPU-enabled counterparts, making it worth to consider investing in heterogeneous computing. Moreover, to the Authors' knowledge, while there have been previous attempts to port MHD codes to GPU's \cite{wong11, wong14, lani14-CPC}, this paper discusses the first and only implicit, time-accurate, multi-fluid plasma code for unstructured meshes to have been ported to GPU's so far. \\
\\
The manuscript is structured as follows: at first, the ideal two-fluid (ion-electron) plasma model from \cite{AlvarezLaguna18-CPC} is recalled; the porting to GPU of the corresponding object-oriented solver is discussed in the second section; the interfacing of the new-generation PARALUTION linear system solving library is presented next; numerical results and benchmarks follow, showing a remarkable improvement in run-time performance up to 14.2x.

\section{COOLFluiD Multi-fluid plasma solver}\label{sec:CFD_solver}

This section describes the parallel implicit, steady/unsteady cell-centered Finite Volume (FV) solver for multi-fluid plasma that has been ported to the GPU. Both the original CPU-enabled version of the code and its GPU-enabled counterpart have been implemented within the COOLFluiD platform\footnote{Inside the COOLFluiD platform, the name given to this solver is MultiFluidMHD.}. 

\subsection{COOLFluiD platform}\label{ssec:CFD_coolfluid}

 The present work has been performed within COOLFluiD  ~\cite{lani1,lani13, Kimpe05}, an open-source software environment for scientific HPC where different numerical techniques, physical models, post-processing algorithms can coexist and work together harmoniously within an underlying highly object-oriented infrastructure. Herein, each numerical method or physical model is encapsulated into an independent dynamic module (or {\it plug-in} library) that can be loaded on demand by user-defined applications. Among its main features, COOLFluiD includes: implicit parallel solvers for compressible/incompressible flows and plasmas using multiple discretization methods for arbitrary unstructured grids (e.g. Finite Volume, Residual Distribution, high-order Flux Reconstruction), interfaces to linear systems solving package (e.g. PETSc, Trilinos, Pardiso), aerothermochemical models for flows \cite{panesi07, degrez09} and plasma \cite{munafo13, panesi13, zhang16}, MHD \cite{yalim06, Yalim11} and radiation transport algorithms based on Monte Carlo \cite{santos16}.

\subsection{Governing equations}\label{ssec:CFD_physics}

In the multi-fluid plasma model, separate conservation equations for mass, momentum and energy are solved for each species, and are coupled to the electromagnetic field computed with the full Maxwell equations. The complete system of PDE's in conservative form reads:

\begin{equation}
\label{eq:ConservativeForm}
\frac{\partial \textbf{U}(\textbf{P})}{\partial t}  + \vec{\nabla} \cdot \vec{ \textbf{F}}^{(c)} = \vec{\nabla} \cdot \vec{\textbf{F}}^{(d)} + \textbf{S}.
\end{equation}
Here, $\textbf{U}$ is the vector of conservative variables and $\textbf{P}$ is the vector of primitive or solution variables. The fluxes are separated in convective, $\vec{\textbf{F}}^{(c)}$, and diffusive, $\vec{\textbf{F}}^{(d)}$, contributions while the source terms are represented by $\textbf{S}$. The full Maxwell's equations add two more constrains for the electric and magnetic fields, namely the Gauss' Law ($\nabla \cdot \vec{E} = \rho/ \epsilon_0$) and the divergence free condition for the magnetic field ($\nabla \cdot \vec{B} = 0 $). Analytically, if both are fulfilled at an initial time, they will be satisfied for all times. However, discretization errors can create a non-zero divergence solution. This problem is tackled in this paper by solving for the evolution equation of the Lagrange multipliers of the Hyperbolic Diverge Cleaning (HDC) method \cite{Munz83}.
For the electromagnetic field, the conservative and primitive variables are equal, and the fluxes can be written as:

\begin{equation}
\label{eq:MaxwellConservative_EM}
\textbf{U} = \textbf{P} =  \begin{pmatrix}
\vec{B} \\
\vec{E} \\ 
\Psi \\
\Phi
\end{pmatrix} , \; \; \;
\vec{\textbf{F}}^{(c)} = \begin{pmatrix}
\bar{ \bar I} \times \vec{E} + \gamma^2 \Psi \bar{\bar I} \\
-c^2 \bar{\bar I} \times \vec{B} + (\chi c)^2 \Phi \bar{\bar I} \\
c^2 \bar{B}^{T} \\
\vec{E}^{T} \\
\end{pmatrix}  , \; \; \;
\textbf{S} = \begin{pmatrix}
0 \\
- \frac{\vec{j}}{\epsilon_0} \\
0 \\
\frac{\rho_c}{\epsilon_0} \\
\end{pmatrix},
\end{equation}
where $\Phi$ and $\Psi$ are the Lagrange multipliers for a divergence-free solution. 

In this work, we have studied an ideal two-fluid plasma model which considers electrons and ions that only interacts with the electromagnetic field. We denote the electron variables with the subindex $e$ and the ion variables with $i$. The electromagnetic field is computed with the full Maxwell equations which take into account charge separation and displacement currents, i.e., $\rho_c = \rho_e q_e + \rho_i q_i$ and $\vec{j} = q_e n_e \vec{u}_e + q_i n_i \vec{u}_i$ respectively. The conservation equations of Eq.~\eqref{eq:ConservativeForm} now reads, separating the electromagnetic and fluid components for clarity:

\begin{equation}
\label{eq:MaxwellConservativeTwoFluid_EM}
\textbf{U}_{EM} = \textbf{P}_{EM} =  \begin{pmatrix}
\vec{B} \\
\vec{E} \\ 
\Psi \\
\Phi
\end{pmatrix} , \; \; \;
\vec{\textbf{F}}_{EM} = \begin{pmatrix}
\bar{ \bar I} \times \vec{E} + \gamma^2 \Psi \bar{\bar I} \\
-c^2 \bar{\bar I} \times \vec{B} + (\chi c)^2 \Phi \bar{\bar I} \\
c^2 \bar{B}^{T} \\
\vec{E}^{T} \\
\end{pmatrix}  ,
\end{equation}
\begin{equation}
\textbf{S}_{EM} = \begin{pmatrix}
0 \\
- \frac{1}{\epsilon_0} \left( q_e n_e \vec{u}_e + q_i n_i \vec{u}_i \right) \\
0 \\
\frac{1}{\epsilon_0} \left( n_e q_e + n_i q_i \right) \\
\end{pmatrix}
\end{equation}

\begin{equation}
\label{eq:MaxwellConservativeTwoFluid_fluids}
\textbf{U}_s = \begin{pmatrix}
\rho_s \\
\rho_s \vec{u}_s \\
\rho_s \varepsilon_s \\
\end{pmatrix}, \; \; \;
\textbf{P}_s =  \begin{pmatrix}
\rho_s \\
\vec{u}_s \\
T_s \\
\end{pmatrix} , \; \; \;
\vec{\textbf{F}}_s = \begin{pmatrix}
\rho_s \vec{u}_s \\
\rho_s \vec{u}_s \otimes \vec{u}_s + p_s \bar{\bar I} \\
\rho_s H_s \vec{u}_s
\end{pmatrix} , 
\end{equation}

\begin{equation}
\label{eq:MaxwellConservativeTwoFluid_source}
\textbf{S}_s = \begin{pmatrix}
0 \\
\rho_s q_s \left[\vec{E} + \vec{u}_s \times \vec{B} \right]  \\
\rho_s q_s \vec{u}_s \cdot \vec{E} \\
\end{pmatrix}, \quad \text{with}\quad s \in \{ e, i \}.
\end{equation}
Here, we simply denote $\vec{\textbf{F}}^{(c)}$ as $\vec{\textbf{F}}$ because in this model the dissipation is neglected ($\vec{\textbf{F}}^{(d)} = 0$). Each species has its own density, momentum and internal energy: $\rho_s$, $\rho_s \vec{u}_s$ and $\rho_s \varepsilon_s$ respectively. The specific internal energy of each species is given by $\varepsilon_s= c_v T_s + u_s^2/2 $, where $T_s$ is the temperature of the specie, and their specific enthalpies are $H_s = \varepsilon_s + p_s/\rho_s$. This model sums a total of 18 variables and their respective partial differential equations.

\subsection{Space discretization}\label{ssec:CFD_space_disc}

The finite volume (FV) discretization that is presented follows the implementation done by Lani et al.~\cite{lani14-CPC}. The scheme provides second-order accuracy thanks to a least-square reconstruction \cite{barth94} of the variables at the faces, based on a stencil including all face-vertex cell neighbors.

In general, the finite volume method (FVM) is based on the integral form of the conservation equations, i.e, Eq.~\eqref{eq:ConservativeForm}:

\begin{equation} \label{eq:FVgeneralMethod}
\frac{d}{dt}\int_{\Omega_i} \textbf{U}(\textbf{P}) d\Omega  = - \oint_{\Sigma_i} \vec{\textbf{F}} \cdot \vec{n} d\Sigma + \int_{\Omega_i} \textbf{S} d\Omega.
\end{equation}
Here, we have divided our computational domain into N non-overlapping volumes $\Omega_i$, each of them enclosed in a surface $\Sigma_i$ with outward normal unit vector $\vec{n}$. We can express the integrals of Eq.~\eqref{eq:FVgeneralMethod} as a function of the cell-averaged values, as follows

\begin{equation} \label{eq:FVdiscretized}
\frac{d}{dt}\textbf{U}(\textbf{P}_i) \Omega_i = - \sum_{j \in \mathcal{D}_i} \mathcal{H} (\textbf{U}_R, \textbf{U}_L, \vec{n}_{ij})\Sigma_{ij} + \textbf{S}_i \Omega_i,
\end{equation}
where $\Sigma_{ij}$ is the interface area between the i-th cell and the neighboring cell $j$ that belongs to the set of neighbors ($\mathcal{D}_i$), and $\textbf{U}_{R/L}$ are the reconstructed values at the left and right sides of the cell interface. The state vector is denoted as $\textbf{P}_i$. The source is averaged over the cell, and the flux is approximated by a function $\mathcal{H}$ that depends on the reconstructed variables at the j-th face. This numerical flux functions must satisfy the consistency property, i.e., $\mathcal{H}(\textbf{U}, \textbf{U}, \vec{n}) = \vec{\textbf{F}} \cdot \vec{n} \equiv \textbf{H} (\textbf{U})$. In the following sections, we will briefly describe the numerical flux function $\mathcal{H}$ for both the electromagnetic and the fluid variables.

\subsubsection{Numerical fluxes discretization}\label{ssec:NumericalFluxes}

For the electromagnetic variables we have used the same approach as in Alvarez Laguna et al.~\cite{AlvarezLaguna18-CPC}, where the Maxwell equations are discretized by a modified Rusanov scheme with scaled dissipation. For those variables, the numerical flux function reads:

\begin{equation}
\mathcal{H}(\textbf{U}_L, \textbf{U}_R, \vec{n}) = \frac{\textbf{H} (\textbf{U}_L) + \textbf{H} (\textbf{U}_R) }{2} - \frac{1}{2} | \textbf{A}_n | \left( \textbf{U}_R - \textbf{U}_L \right),
\end{equation}
where $|\textbf{A}_n|$ is the numerical dissipation, whose detailed expression can be found in Alvarez Laguna et al.~\cite{AlvarezLaguna18-CPC}.

The fluid variables are discretized with a generalization of the AUSM$^+$-up scheme for all speeds~\cite{Liou06} (subsonic, transonic and supersonic regimes) for the multi-fluids equations ~\cite{AlvarezLaguna18-CPC,AlvarezLaguna16}, that reads

\begin{equation}
\textbf{H}_s (\textbf{U}) = M_s a_s 
\begin{pmatrix}
\rho_s \\ \rho_s \vec{u}_s \\ \rho_s H_s
\end{pmatrix}
 + p_s
 \begin{pmatrix}
 0 \\ \vec{n} \\ 0
 \end{pmatrix},
\end{equation}
where expressions for the $M_s$, $a_s$ and $p_s$ can be found in ~\cite{AlvarezLaguna16,Liou06}.

\subsubsection{Implicit time integration}\label{ssec:CFD_time_disc}

Once the spatial discretization is performed, the array of residuals is defined from Eq.~\eqref{eq:FVdiscretized} as:

\begin{align}
\textbf{R}(\textbf{P}) &= [\textbf{R}_0(\textbf{P}_0), \cdots, \textbf{R}_{N-1}(\textbf{P}_{N-1})] \nonumber \\ 
&\text{with}  \quad \textbf{R}_i(\textbf{P}_i) = - \sum_{j \in \mathcal{D}_i} \mathcal{H}(\textbf{U}_L, \textbf{U}_R, \vec{n}_{ij})\Sigma_{ij} + \textbf{S}_i \Omega_i.
\end{align}

Similarly, the \textit{pseudo-steady} residuals are defined as
\begin{eqnarray}
  {\tilde {\bf R}}({\bf P}) = \dd{{\bf U}}{{\bf P}} 
  \dd{{\bf P}}{t} + {\bf R} ({\bf P}) = {\bf 0},
\end{eqnarray}

For the present work, the study will be focused on unsteady cases. In order to have a stable implicit discretization, we discretize ${\tilde {\bf R}}({\bf P})$ as:
\begin{eqnarray}
  {\tilde {\bf R}} ({\bf P}) & = & \frac{3 {\bf U}({\bf P}) - 
    4 {\bf U}({\bf P}^k) + {\bf U}({\bf P}^{k-1})}{2 \Delta t} \Omega + {\bf R}
  ({\bf P}),
\end{eqnarray} 
which corresponds to a 3-point backward Euler algorithm, second-order accurate in time \cite{Butcher2016}.

In order to solve the implicit system, a Newton linearization is performed at each sub-iteration step, yielding
\beq
\left \{ 
  \begin{array}{rcl} 
    \left[ \dd {{\bf {\tilde R}}}{{\bf P}} \left({\bf P}^k
      \right) \right] \; 
    \Delta {\bf P}^k &=& - {\bf {\tilde R}}({\bf P}^k) \\
    
    \\
    
    {\bf P}^{k+1} &=& {\bf P}^{k} + \Delta {\bf P}^k
  \end{array} \right.,
\label{eq:newton_system}
\eeq
where $\dd {{\tilde {\bf R}}}{\bf P}$ is the Jacobian matrix. For those unsteady cases, we keep iterating until $\parallel\Delta {\bf P}^k\parallel < \varepsilon$  and then set ${\bf P}^{n+1} = {\bf P}^{k_{last}+1}$.

The Jacobian is computed numerically \cite{phd:lani08} for both the convective and source terms. The analytical Jacobian for the multi-fluid AUSM$^+$-up plus the modified Rusanov scheme is rather complex to implement, therefore we have preferred to rely on a less error-prone and flexible (that can work with any model/scheme out of the box) approach, at a cost of losing some performance. In principle, a Jacobian-Free Newton Krylov (JFNK) approach, not requiring the storage of the system matrix, could be used but would require considerable additional development which would impact not just the Finite Volume implementation but also the COOLFluiD interface to the PARALUTION and PETSc libraries. This is definitely out of the scope of the present paper, but it could be part of future work to evaluate the impact of JFNK on the performance of the target applications and could be advantageous to save a considerable amount of memory, as long as also the preconditioner remains matrix-free (Jacobi or Block-jacobi) as well.

A more detailed description of the previous numerical scheme, as well as an extensive validation, can be found in the following references, \cite{AlvarezLaguna16} for the multi-fluid model with the quasi-neutral assumption, and \cite{AlvarezLaguna18-CPC} for the ideal ion-electron model with charge separation.

\subsubsection{Linear system solvers}

During this work, two different Linear System Solvers (LSS) were used to solve the first step of Equation \eqref{eq:newton_system}:

\begin{itemize}
\item PETSc \cite{petsc-web-page}, which is an open-source library that is extensively used in large-scale projects and can also solve linear systems by taking advantage of GPU's \cite{petsc-GPU}.
\item PARALUTION \cite{paralution}, which is a library devoted to solving sparse systems through different algorithms and preconditioners. It is based on C++ and supports multi-node CPU's and GPU's systems. In its single-node version\footnote{The multi-node version of PARALUTION is under a commercial license.}, the computation can be accelerated using only one GPU. This is the only case studied in this work. It has the advantage that the matrix can be allocated and updated directly on the GPU.
\end{itemize}

The COOLFluiD interface with PETSc was already implemented and tested \cite{lani14-CPC}. On the other hand, the interface with PARALUTION is new and will be discussed afterwards.

\section{Porting to GPU}\label{sec:GPU}

The work carried out in this paper is an extension to the parallel implementation developed in COOLFluiD and used for MHD simulations \cite{lani14-CPC}. The tools and concepts that were developed by Lani et al.~\cite{lani14-CPC} are thus extensively used throughout this work. Nevertheless, the necessary requirements and structure for a numerical method to be ported to GPU within COOLFluiD and the parallel computation of the residual and its Jacobian will be discussed.

\begin{figure}[!ht]
   \centering
   \includegraphics[width=0.5\textwidth]{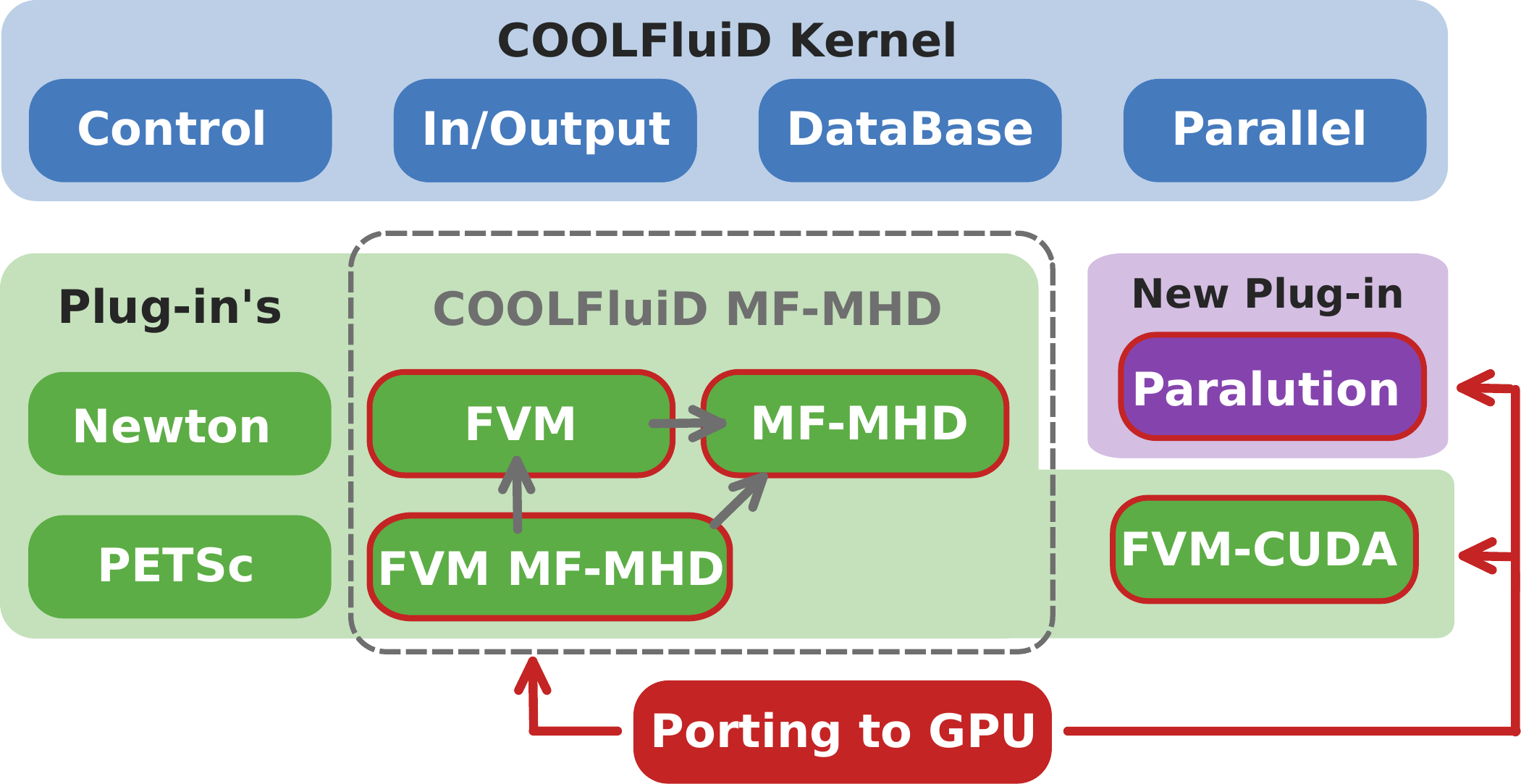}
   \caption{Schematics of the COOLFluiD MultiFluidMHD in the current version. The porting to GPU involved previously existing modules (e.g., FVM MF-MHD) and the new implementation of a PARALUTION wrapper module.}\label{fig:CF_GPU}
 \end{figure}

The code design used in COOLFluiD is sketched in Fig.~\ref{fig:CF_GPU}. The kernel provides the high-level control of the simulation, parallel data-structure and I/O capabilities, a set of basic interfaces for models and algorithms. The core of the COOLFluiD multi-fluid plasma solver includes:

\begin{itemize}
\item 
  the MultiFluidMHD module featuring the governing equations (convective fluxes, variable sets, transformations);
\item 
  the FVM module implementing a generic FV algorithm;
\item
  the FVM-CUDA module which adapts the FVM to compute the residual and Jacobian on the GPU;
\item 
  the FVM-MutiFluidMHD module collecting some FV schemes and boundary conditions tailored to MultiFluidMHD.
\end{itemize}
Other useful libraries which are dynamically linked to the main solver are:
\begin{itemize}
\item
  the Newton iterator module implementing implicit time integrators;
\item  
  the interface to the PETSc toolkit which is used to solve linear systems.
\item
  the newly created interface for the PARALUTION library for solving linear systems.
\end{itemize}

\subsection{GPU-enabled model}

In a general case, the \textit{model} stores all the information regarding the physics of the system. As COOLFluiD is a multi-physics platform, the set of equations to solve and the variables to describe the systems can change drastically. For example, a \textit{model} has to have the capability of computing the flux, given certain variables, i.e., $\vec{\textbf{F}}$, and should store all the physical constants of the system, e.g., the speed of light.

\begin{figure}[!h]
  \centering
  \includegraphics[width=0.5\textwidth]{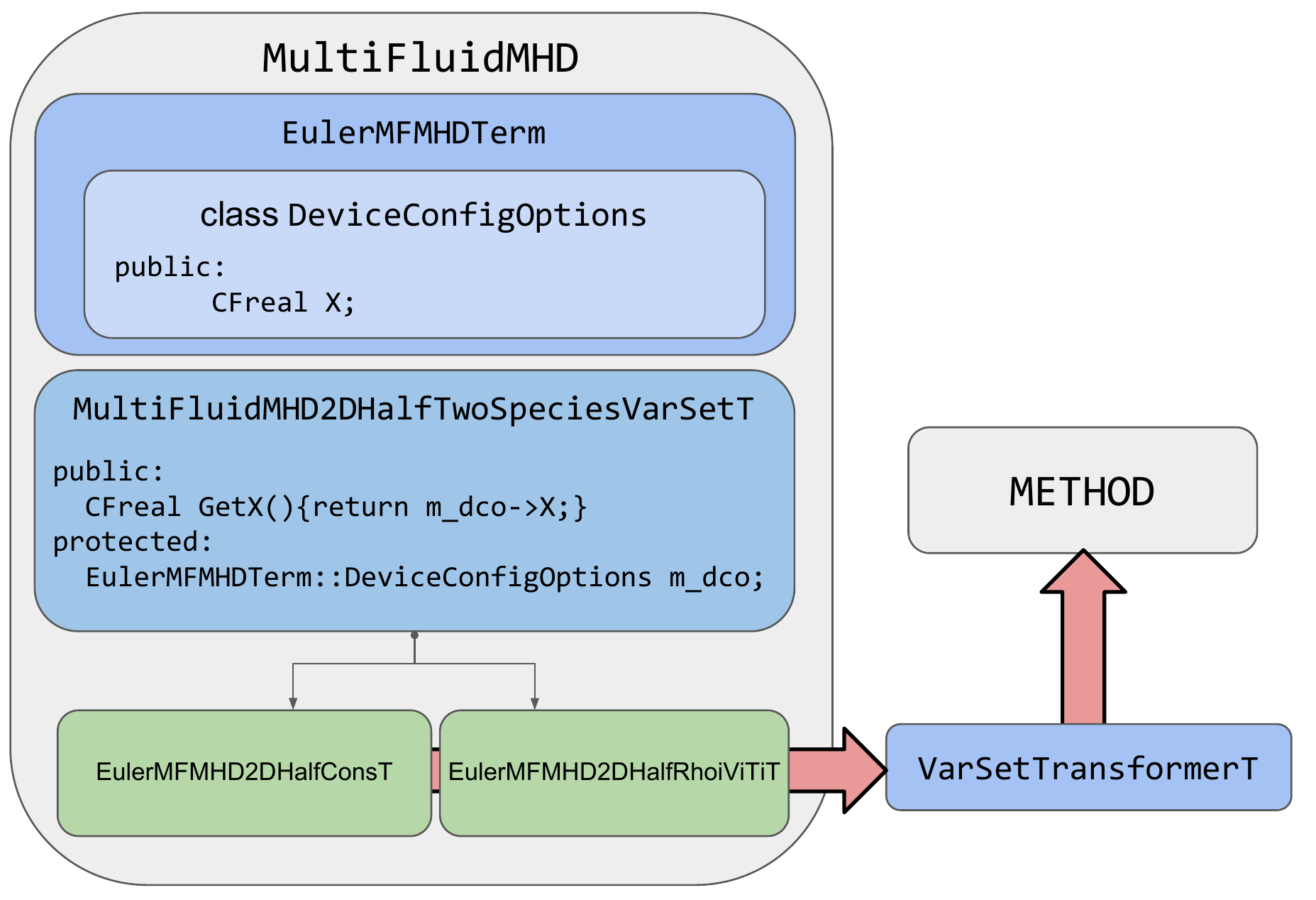}
  \caption{Structure of the GPU-enabled multi-fluid \textit{model}. All the information about the physics and equations are inside the MultiFluidMHD module. A variables set (such as the conservative variables) can get information from the physical terms via a \texttt{dco} object. These variable sets are passed as template parameters to a generic variable transformation object, which is a template parameter of a generic method (e.g. a numerical flux function).}
  \label{fig:ModelTemplate}
\end{figure}

The structure of a GPU-enabled model changes with respect to the CPU implementation, as all the parameters needed in the computation should be copied to the GPU memory during run-time. The structure developed is shown in Fig.~\ref{fig:ModelTemplate}. 

\begin{itemize}
\item The core of the multi-fluid physical model is the \texttt{EulerMFMHDTerm} class, which stores configurable values that are given by the user (e.g., the specific heat ratio). Nested to it, the \texttt{DeviceConfigOption} (\texttt{dco}) class is created, as well as two \texttt{copy()} functions. The first one is in charge of copying the data to the \texttt{dco} (stored on the GPU), and is only called if the user adds the flag \texttt{onGPU=true} in the configuration file, the second one copies the parameters to the \texttt{dco} when computing on the CPU.
\item \texttt{MultiFluidMHD2DHalfTwoSpeciesVarSetT} is the base class for the different variables set, e.g., conservative and primitives. As a protected value, it has the \texttt{dco} of \texttt{EulerMFMHDTerm}. In order to read the parameters that are stored in it, \texttt{get()} functions are created.
\item For each variable set a different child class is created. Usually, one for conservative variables: \texttt{EulerMFMHD2DHalfConsT}; and another one for the update variables (i.e., the variables in which the solution is stored and updated): \texttt{EulerMFMHD2DHalfRhoiViTiT}.
\item To transform from each one of the variable sets, a \texttt{VarSetTransformerT} object is created, which accepts as template parameters two variable sets. Then a \texttt{transform()} function computes the change of variables between the sets.
\item Finally, the \texttt{VarSetTransformerT} is passed to a generic GPU-enabled method as a template parameter, so all the variables that are stored in the \texttt{dco} of the physical model are accessible, as well as the variable sets and how to transform them.
\end{itemize}

\subsection{GPU-enabled method}

Fig.~\ref{fig:MethodTemplate} presents the general structure of the implementation of a GPU-enabled {\it method}. We use this structure for methods that compute some numerical algorithms that do not explicitly depend on the physics. A clear example of these methods is the numerical flux function, $\mathcal{H}$, whose implementation does not depend upon the physics but can take certain outputs of a model to do a computation, e.g., $\textbf{H} (\textbf{U}_{L/R})$. The source terms are also defined as \textit{methods} to ease the implementation, as they can be easily turned on/off without changing the \textit{model}. 

Inside the \textit{method} class, two nested objects are created to allow the computations to be carried out at the GPU (following the overall structure described at \cite{lani14-CPC}): 
\begin{itemize}
\item The class \texttt{DeviceConfigOptions} (\texttt{dco}) stores in the GPU, or in the CPU if the user-defined option \texttt{onGPU=false}, the parameters that are needed for the method to work. In the same fashion than the model, these are given by the user in the configuration file. The \texttt{init()} function initializes the parameters as a copy of another object of the same kind. And finally, two different functions are created to copy the parameters to the \texttt{dco}.

\begin{figure}[!h]
  \centering
  \includegraphics[width=0.5\textwidth]{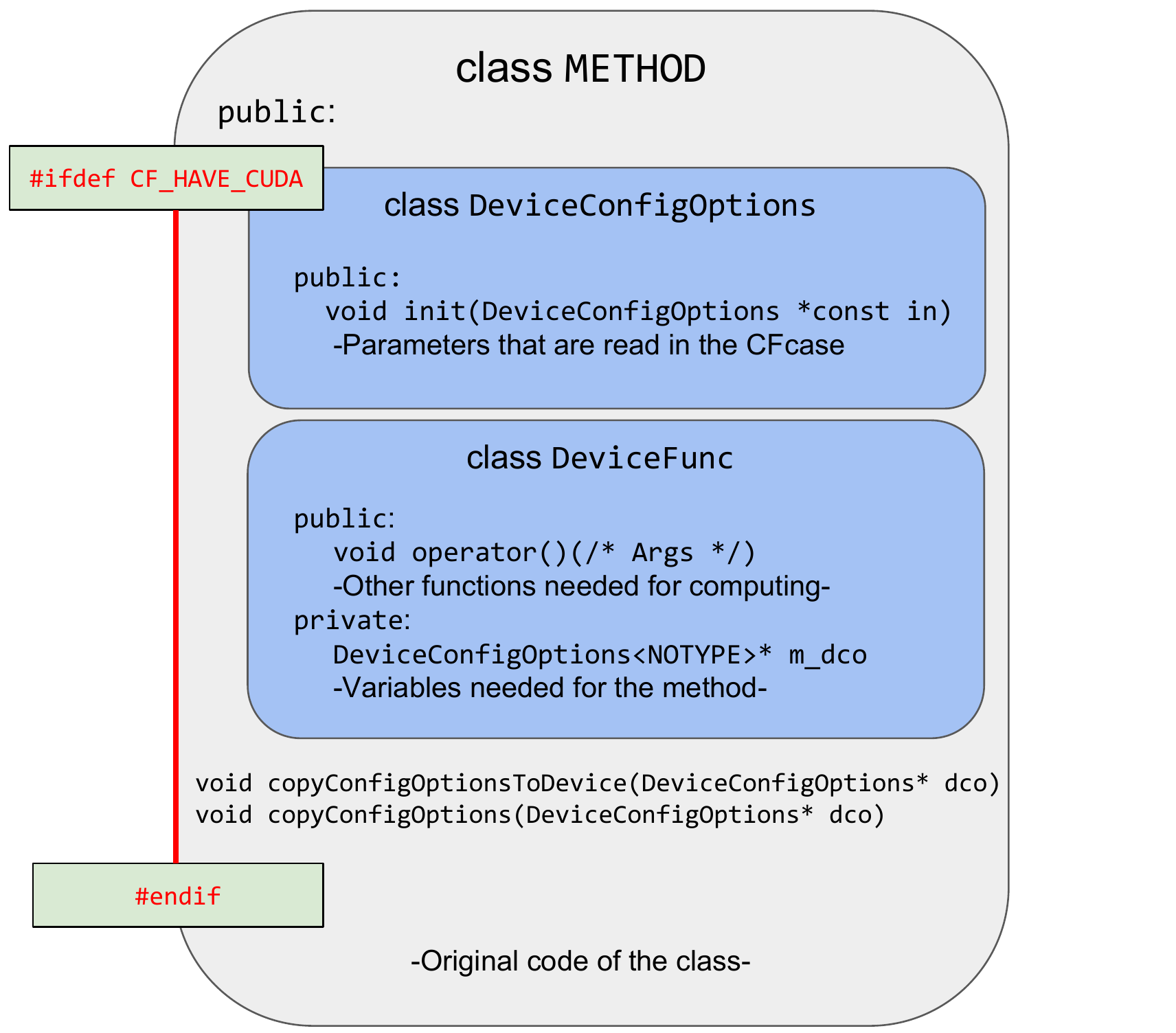}
  \caption{General GPU-enabled \textit{METHOD} structure. Two nested objects are defined, one for storing configuration parameters and another one for enclosing the actual GPU implementation of the method.}
  \label{fig:MethodTemplate}
\end{figure}

\item The class \texttt{DeviceFunc} is a nested functor, i.e., the \texttt{operator()} is overloaded and can be called like a function. This object is called from the computation of the residual and its Jacobian, and it is the core of the implementation of the method. Inside \texttt{operator()}, there is essentially the same algorithm as in previous works, but it is adapted to run on the GPU. 
\end{itemize}

Although the COOLFluiD platform allows and takes full advantage of dynamic allocation of arrays, for the GPU port we have chosen to statically allocate the needed arrays. This way we avoid dynamic allocation of arrays on the GPU, which in general is expensive and can damage the performance \cite{CudaHandbook}.

In this work, three different methods have been ported to run on the GPU:

\begin{enumerate}
\item AUSM+Up / Modified Rusanov schemes for computing the numerical flux function at the cell interfaces.
\item Venkatakrishnan limiter \cite{Venktn93}. Previously, only a Barth-Jespersen limiter was ported to run on GPU. However, as illustrated in Alvarez Laguna et al.~\cite{AlvarezLaguna18-CPC}, the limiter is of vital importance on multi-fluids plasma simulations and consequently the porting to GPU was needed to obtain the same results.
\item Ideal two-fluid source term, Eq. \eqref{eq:MaxwellConservativeTwoFluid_source}, which couples the fluids to the electromagnetic field via the Lorentz force.
\end{enumerate}

\subsection{Parallel flux-source computation}

In the previously existing MHD GPU-enabled Finite Volume algorithm, the source term was not implemented. This is due to the fact that Ideal MHD equations can be written in conservation form, i.e., the Lorentz force can be written as a flux. However, in the multi-fluid model, the Lorentz force that is exerted on each fluid as well as the source terms in Maxwell's equations cannot be written as a flux. Consequently, the implementation of the source terms in the GPU-enabled multi-fluid FVM solver is mandatory. In order to compute these terms, the FVM-CUDA module was modified and new classes were created.

Previously, the numerical flux function, limiter, and other methods were given as template parameters to the objects that compute the residual, $\textbf{R}(\textbf{P})$ , and its Jacobian ,$\partial \textbf{R}/\partial \textbf{P}$, e.g., \texttt{ComputeRhsJacobCell}. However, in order to add one more template parameter, a new object should be created that can accept one source term, and only one. This can be changed in the future to allow for multiple source terms within the same object. 

These new objects, e.g. \texttt{ComputeSourceRhsJacobCell}, share part of the previous implementation. Nevertheless, the kernel call that previously had computed the flux and its Jacobian, now computes the flux, source term and their Jacobian, as seen in algorithm \ref{alg:computeFluxSourceJacobian}.

\begin{algorithm}
\caption{Kernel to compute the residual and its Jacobian.}
\label{alg:computeFluxSourceJacobian}
\begin{algorithmic}
\Function{computeFluxSourceJacobianKernel}{}
	\State CellID = threadIdx.x + blockIdx.x*blockDim.x + startCellID;
	\If {CellID $<$ nbCells}
    	\State Initialize objects
        \State set cellData
    	\State Reset residual
    	\For{ face in nbFacesInCell }
    		\State Set faceData
    		\State \Call{extrapolateOnFace}{cellData, faceData}
            \State \Call{fluxScheme}{faceData, model}
            \State Update residual
            \For{variable in NbEqs}
            	\State Perturb variable of the left cell
                \State \Call{extrapolateOnFace}{cellData, faceData}
                \State \Call{fluxScheme}{faceData, model}
                \State Compute residual derivative
                \State Add contribution to diagonal block
                \State Add contribution to off-diagonal block
                \State Restore variable
            \EndFor
	   	\EndFor
        \State \Call{Source} {cellData, model}
        \State Update residual
        \For{variable in NbEqs}
        	\State Perturb variable
            \State \Call{Source} {cellData, model}
            \State Compute residual derivative
            \State Add contribution to diagonal block
            \State Restore variable
        \EndFor
	\EndIf
\EndFunction
\end{algorithmic}
\end{algorithm}

Each GPU thread is mapped to a cell, and the first step is to initialize all the needed objects from the \texttt{dco}'s. These objects include the physical model, source term, numerical flux function, limiter and spatial discretization method. The first loop is done over the cell faces, where the flux is computed with the reconstructed states at the face. Within the same loop, the derivatives are computed and added to the corresponding entries of the Jacobian matrix, linking both cells. When this loop finishes, the source term is computed and added to the residual, and afterwards its Jacobian is computed by perturbing the different variables one by one and computing the corresponding source term contributions.

The boundary conditions are always computed in the CPU, as in regular cases the ratio of inner cells to boundary cells is high enough to neglect the increase in efficiency that can be achieved by porting the boundary computations to the GPU.

\section{PARALUTION interface} \label{sec:paralution}
In this section, we present the interface with the linear solver library PARALUTION on GPU's and briefly compare to PETSc as in Section \ref{sec:Results} the performance of both libraries will be compared. PETSc ~\cite{petsc-web-page, petsc-user-ref, petsc-efficient} is a widely used library for solving linear systems of equations on the CPU. It offers a large set of solvers and preconditioner that are optimized for serial and parallel execution. The GPU version of the library, however, does not provide a straightforward way to integrate arrays computed on the GPU into a system matrix. Consequently, the assembled data on the GPU is passed to PETSc via the CPU. In order to solve the system of equations on the GPU, an additional transfer of the system matrix and the residual to the GPU is needed. This is suboptimal, and the efficiency can decrease noticeably for fine meshes. In order to tackle this problem and further improve the performance of the code, a new option is studied, which consists of developing a wrapper module to PARALUTION.

Unlike PETSc, the PARALUTION library offers the possibility to build and allocate the linear system directly on the GPU, thus avoiding the time-consuming copying back and forth the Jacobian and residual. In order to take advantage of this, the FVM-CUDA module was modified to add a new kernel, which computes the spatial discretization Jacobian and, within the same call, stores the computed values in its corresponding entries at the matrix. This kernel follows the same algorithm specified before, and the only difference is the way in which the Jacobian contribution is added to the matrix. Using PETSc, this is done in blocks, whereas with PARALUTION they are added directly to the arrays in compressed sparse row (CSR) format.

\begin{figure}[!h]
  \centering
  \includegraphics[width=0.5\textwidth]{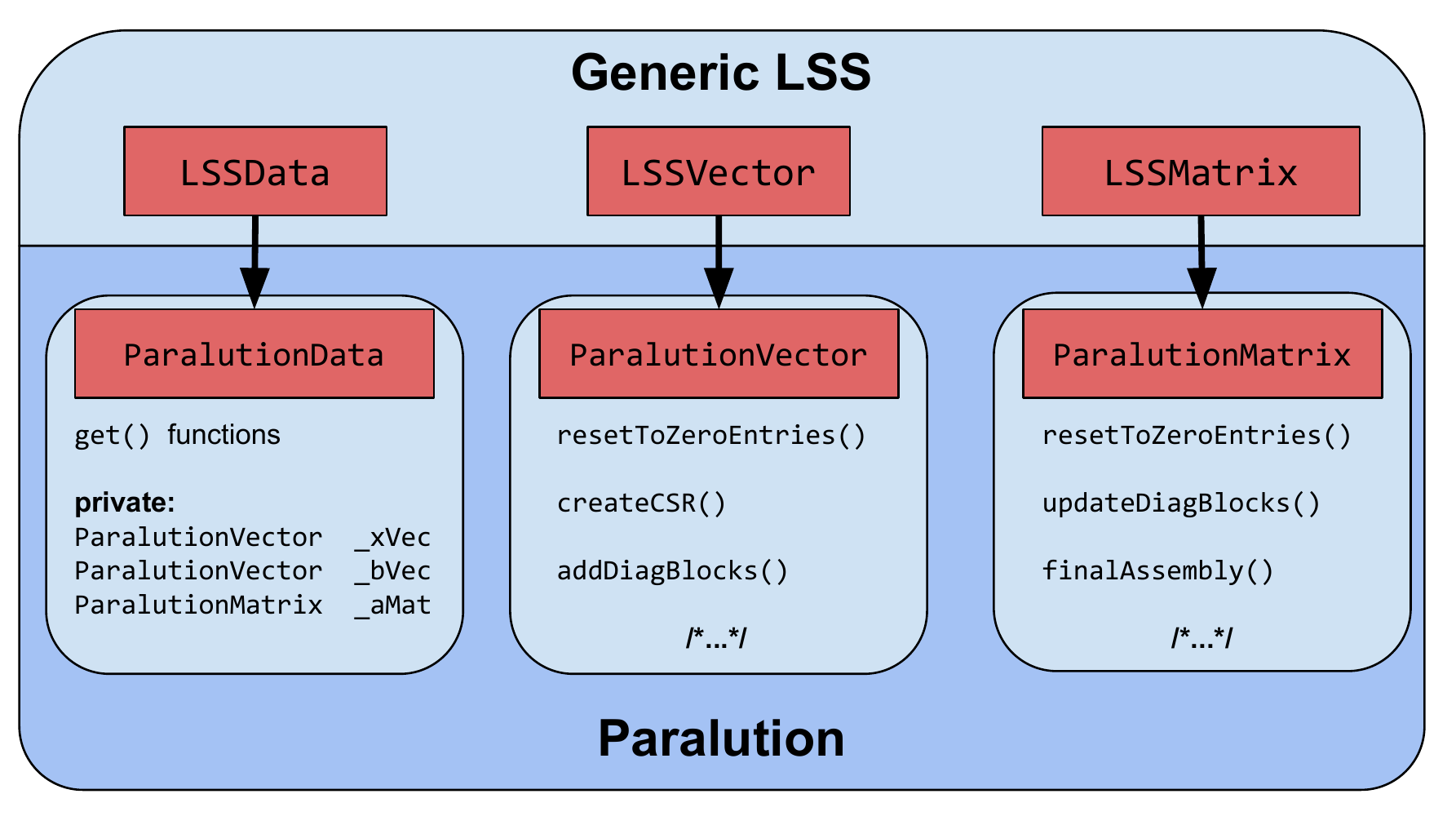}
  \caption{Structure of PARALUTION data objects.}
  \label{fig:ParalutionModelTemplate}
\end{figure}

The core of this newly implemented module is similar to the existing PETSc wrapper module, which  is enclosed in general LSS objects in COOLFluiD (Fig. \ref{fig:ParalutionModelTemplate}). 

A boolean parameter was defined inside the PARALUTION module, which determines whether the system is directly saved on the GPU memory, or on the RAM. The newly developed functions were designed to access both GPU or CPU memory. For example, during the setup phase, the needed memory for the system, in CSR format, is allocated in the CPU, and in the GPU if needed (code listing \ref{lst:createCSR}).


\begin{lstlisting}[caption={ParalutionMatrix::createCSR() implementation},label={lst:createCSR}]
// ------ ParalutionMatrix.cxx ------ //
void ParalutionMatrix::createCSR(std::valarray<CFint> allNonZero, CFuint nbEqs, CFuint nbCells)
{
   
   CFuint nbStates = allNonZero.size();
   _rowlength = nbStates+1;

   _rowoff = new CFint[_rowlength];  
   _size = 0;
   _rowoff[0] = 0;

   for (CFuint i=0; i<nbStates; i++){ 
      _rowoff[i+1] = _size+allNonZero[i];
      _size += allNonZero[i];
   }

   _val = new CFreal[_size];   
   _col = new CFint[_size];   
   
  initializeMatrix();  //Set the values to zero
  if(buildOnGPU){
      
    _nbKernelBlocks = 64;
    _blocksPerGrid = CudaEnv::CudaDeviceManager::getInstance().getBlocksPerGrid(nbCells);
    _nThreads = CudaEnv::CudaDeviceManager::getInstance().getNThreads();
    _sizeb = (_blocksPerGrid+(_nbKernelBlocks-1))/_nbKernelBlocks;

      CudaEnv::allocDev(_rowoffDev, _rowlength);
      CudaEnv::allocDev(_colDev, _size);
      CudaEnv::allocDev(_valDev, _size);
      CudaEnv::allocDev(_diagAccDev, _nThreads*_nbKernelBlocks*nbEqs*nbEqs);   //KernelSize
      diagAccSize = nbStates*nbEqs*nbEqs;
      _diagAcc = new CFreal[diagAccSize];

      //memcopy()
      CudaEnv::copyHost2Dev(_rowoffDev, _rowoff, _rowlength);
      CudaEnv::copyHost2Dev(_colDev, _col, _size);
      CudaEnv::copyHost2Dev(_valDev, _val, _size);

      m_mat.MoveToAccelerator();
   }
}
\end{lstlisting}

The number of nonzero values for each cell is stored in \texttt{allNonZero}. For instance, if there are 16 equations per cell and the cell has 4 neighbors, the value of \texttt{allNonZero} is $16 \cdot 4 = 64$.
The arrays \texttt{\_valDev} and \texttt{\_colDev} are filled by the GPU kernels responsible of the computation of the residual and its Jacobian. These kernels add, usually, blocks to the matrix, most of them being off-diagonal \footnote{Off-diagonal in the sense that they are linking two different cells.}. Nevertheless, the  contribution to the Jacobian of the time discretization (computed on the CPU) is always on the diagonal block. 
Knowing this beforehand allows for a more efficient algorithm: at each iteration, after the Jacobian of the spatial discretization has been added to the matrix directly on the GPU memory, the time discretization Jacobian is computed on the CPU, where only the diagonal blocks of the Jacobian are allocated. Just before solving the system, these blocks are copied to the GPU and added to the spatial discretization Jacobian (code listing \ref{lst:addDiagBlockGPU}).

One drawback of this method is that the diagonal blocks need to be copied to the GPU, thus requiring extra GPU memory. Due to the relatively limited device memory, this lowers the maximum system size which is able to run on GPU. To lessen the impact on the maximum system size, the diagonal blocks are copied in chunks to the GPU. Therefore, the memory usage is decreased at the expense of increasing the computational time.


\begin{lstlisting}[caption={ParalutionMatrix::addDiagBlockGPU() implementation},label={lst:addDiagBlockGPU}]
// ------ ParalutionMatrix.cxx ------ //
__global__ void ParalutionMatrix::addDiagBlockGPU(CFreal *val, CFint *rowoff, CFreal *accDevPtr, CFuint nb, CFuint startCellID, CFint nbCells)
{
  const int cellID = threadIdx.x + blockIdx.x*blockDim.x + startCellID;
  if (cellID < nbCells) { 
    CFuint RowPositionDiag = rowoff[cellID*nb];
    CFuint RowPositionPlusOneDiag = rowoff[cellID*nb + 1];
    CFuint mmDiag = (RowPositionPlusOneDiag-RowPositionDiag)/nb;
    for(CFint nbi=0; nbi<nb; nbi++){
      for(CFint nbj=0; nbj<nb; nbj++){
          val[RowPositionDiag+nbi*nb*mmDiag+nbj] += accDevPtr[localID*nb*nb + nbi*nb + nbj];
      }
    }
  }
}
\end{lstlisting}


Before solving the system, these device arrays  are directly passed to PARALUTION through the library function \texttt{SetDataPtrCSR}, which has practically no overhead on overtaking the content of the device arrays into the GPU PARALUTION matrix (code listing \ref{lst:finalAssembly}).


\begin{lstlisting}[caption={ParalutionMatrix::finalAssembly() implementation},label={lst:finalAssembly}]
// ------ ParalutionMatrix.cxx ------ //
void ParalutionMatrix::finalAssembly(CFuint size){
  if(buildOnGPU){
    CFLog(VERBOSE, "ParalutionMatrix::finalAssembly -> GPU \n");
    m_mat.SetDataPtrCSR(&_rowoffDev, &_colDev, &_valDev, "SystemMatrix", _size, size, size);
  }else{
    CFLog(VERBOSE, "ParalutionMatrix::finalAssembly -> CPU \n");
    m_mat.SetDataPtrCSR(&_rowoff, &_col, &_val, "SystemMatrix", _size, size, size);
  }
  firstIter=false;
}
\end{lstlisting}


Finally, after solving the system, the matrix entries are initialized to zero. For this, a simple function is used, shown at code listing \ref{lst:resetToZeroEntriesGPU}.


\begin{lstlisting}[caption={ParalutionMatrix::resetToZeroEntriesGPU() implementation},label={lst:resetToZeroEntriesGPU}]
// ------ ParalutionMatrix.cxx ------ //
void ParalutionMatrix::resetToZeroEntriesGPU(){
  cudaMemset(_valDev, 0.0, _size*sizeof(CFreal));  //Reset the whole matrix at the GPU
  std::fill_n(_diagAcc, diagAccSize, 0);           //Reset the diagonal blocks at the CPU
}
\end{lstlisting}


\section{Numerical results and benchmarks}\label{sec:Results}

In this section the correctness of the implementation will be tested and the speedup will be assessed. The performance of the GPU-enabled code has been tested with different configurations  of CPU's/GPU's as listed  below:

\begin{enumerate}[(a)]
\item \textbf{PETSc CPU - CPU},
  original FVM algorithm on the CPU (i.e. the code used in ~\cite{AlvarezLaguna16,maneva17,AlvarezLaguna18-CPC}), with the system assembly implemented as a single loop over all faces and the linear system is solved using PETSc on the CPU.
\item \textbf{PETSc CPU - GPU},
  new implementation using two loops, one on the GPU over the interior cells and one on the CPU over boundary faces. The residual and the spatial discretization Jacobian assembled on the GPU are copied back to be solved by PETSc on the CPU.
\item \textbf{PETSc GPU - GPU},
  same implementation as the preceding configuration, but the system is solved by PETSc on the GPU.
\item \textbf{PARALUTION GPU},
  new implementation using two loops, one on the GPU over the interior cells and one on the CPU over boundary faces. Both the residual and the spatial discretization Jacobian are directly assembled and solved on the GPU by PARALUTION.
\end{enumerate}

For both linear system solvers the Jacobi preconditioner was used. In all cases the three configurations are tested using one CPU core and one GPU on three different systems (technical information about the GPU's can be found in Table \ref{tab:GPUs}):

\begin{enumerate}
\item \textbf{K10}. Dual 6-core Intel(R) Xeon(R) CPU E5-2640 (15MB cache), 2.50 GHz with 132 Gb of RAM, featuring four NVIDIA Tesla K10 GPU accelerators featuring each two GK104 GPU's (i.e. 8 GPU's in total). 
\item \textbf{K40}. Quad 12-core Intel(R) Xeon(R) CPU E5-2690 v3 (30MB cache) @ 2.60GHz 252 Gb of RAM, featuring 2xTesla K40 GPU accelerators (each 12Gb local memory).
\item \textbf{P100}. Dual 18-core Intel(R) Xeon(R) CPU 6140 (25.75MB cache) @ 2.30 GHz 192 Gb of RAM, featuring four Tesla P100 accelerators (each 16 Gb of local memory).
\end{enumerate}

\begin{table*}[h]
\centering 
\begin{tabular}{c|c|c|c|c|c|}
\cline{2-6}
                           & CUDA cores & Frequency & GFLOPS (double) & Memory & Bandwidth \\ \hline
\multicolumn{1}{|c|}{K10}  & 2x1536*      & 745 Mhz   & 190            & 8 GB  & 2x160* GB/s  \\
\multicolumn{1}{|c|}{K40}  & 2880       & 745 Mhz   & 1430            & 12 GB  & 288 GB/s  \\
\multicolumn{1}{|c|}{P100} & 3584       & 1126 Mhz  & 4670            & 16 GB  & 720 GB/s  \\ \hline
\end{tabular}
\caption{Technical data of the three different GPU's used during this work.*The K10 is formed by two GK104s GPU's.}
\label{tab:GPUs}
\end{table*}

For all systems, the GNU Compiler Collection (GCC) was used, the two first systems using version 4.8.3 and the last one using version 6.4.0. In all cases we set \texttt{-O3} optimization level. The LSS versions used: PETSc 3.9.0 and PARALUTION 1.1.0.

 In the following, the solution for circularly polarized waves will be studied and compared to analytical solutions. Afterwards, the more computationally demanding Geospace Environment Modeling (GEM) challenge \cite{Birn01-GEM} will be run and compared with other works.  Regarding the performance, we will compare different configurations using two metrics: the first one separates the time spent per iteration assembling and solving the system, and the second one  compares the total wall time taken for different configurations, in order to consider the runtime improvement in a production-like environment.

\subsection{Circularly polarized wave}

It is common to use the propagation of Alfvén waves as a standard benchmark for MHD codes as the analytical solution is known. These waves propagate along the background magnetic field $B_0$, and are perturbations to the perpendicular velocity and magnetic field. The exact solution for these waves can be obtained for this ion-electron model, thus we can easily asses if the GPU-enabled implicit solver yields correct results. Given the preceding conditions, the dispersion relation for the ion-electron model reads \cite{AlvarezLaguna18-CPC}:

\begin{equation}
1+ \left( \frac{\omega_{pi}}{k c} \right)^2 \frac{\omega}{\omega + \Omega_{ci}} + \left( \frac{\omega_{pe}}{kc} \right)^2 \frac{\omega}{\omega + \Omega_{ce}} - \frac{\omega^2}{k^2 c^2} = 0,
\end{equation}
where $\omega_{ps}$, $\Omega_{cs}$ are, respectively, the plasma and cyclotron frequencies of the species $s\in\{e,i\}$. If we assume that the perturbation is propagated along a magnetic line in the x-direction we obtain the following eigenvectors:

\begin{align}
\vec{B} = (B_0,~\eta B_0 \cos(kx -\omega t),~\eta B_0 \sin (kx - \omega t))  \\
\vec{u}_s = (0,~\eta V_s \cos(kx - \omega t),~\eta V_s \sin(kx - \omega t)),
\end{align}
where
\begin{equation}
V_s = \frac{\Omega_{cs}}{\omega + \Omega_{cs}} \frac{\omega}{k}.
\end{equation}

The parameter $\eta$ controls the amplitude of the perturbation and is set to $\eta=0.1$ to avoid non-linear effects.

\subsubsection{Numerical results}

The simulations are run in a rectangular domain $-L<x<L$, $-L/2<y<L/2$ with periodic boundary conditions in all sides. We replicate the set-up found at ~\cite{AlvarezLaguna18-CPC, Amano15}, but in our case we will only study the rotated configuration, where the wave defined by $(k_x, k_y)=(\pi/L, 2\pi/L)$ travels with an angle of $\tan^{-1}(0.5)$  with respect to the x-axis. We set the mass ratio $m_i/m_e = 256$, a reduced light speed of $c=10V_i$ and $\beta = p_{tot}2 \mu_0/B_0^2 = 0.1$.

We set the size of our domain, $L$, imposing that $L/\lambda_i = \pi$, being $\lambda_i=c/\omega_{pi}$ the ion inertial length. The mesh is formed by $2N\times N$ cells and tests were carried out with $N=128, 64$ and setting $CFL=1$. Both simulations were run using the PARALUTION - GPU configuration on the K10 system, and were run until the wave had traveled five wavelengths.

\begin{figure*}[ht!]
\centering
\includegraphics[width=0.7\textwidth]{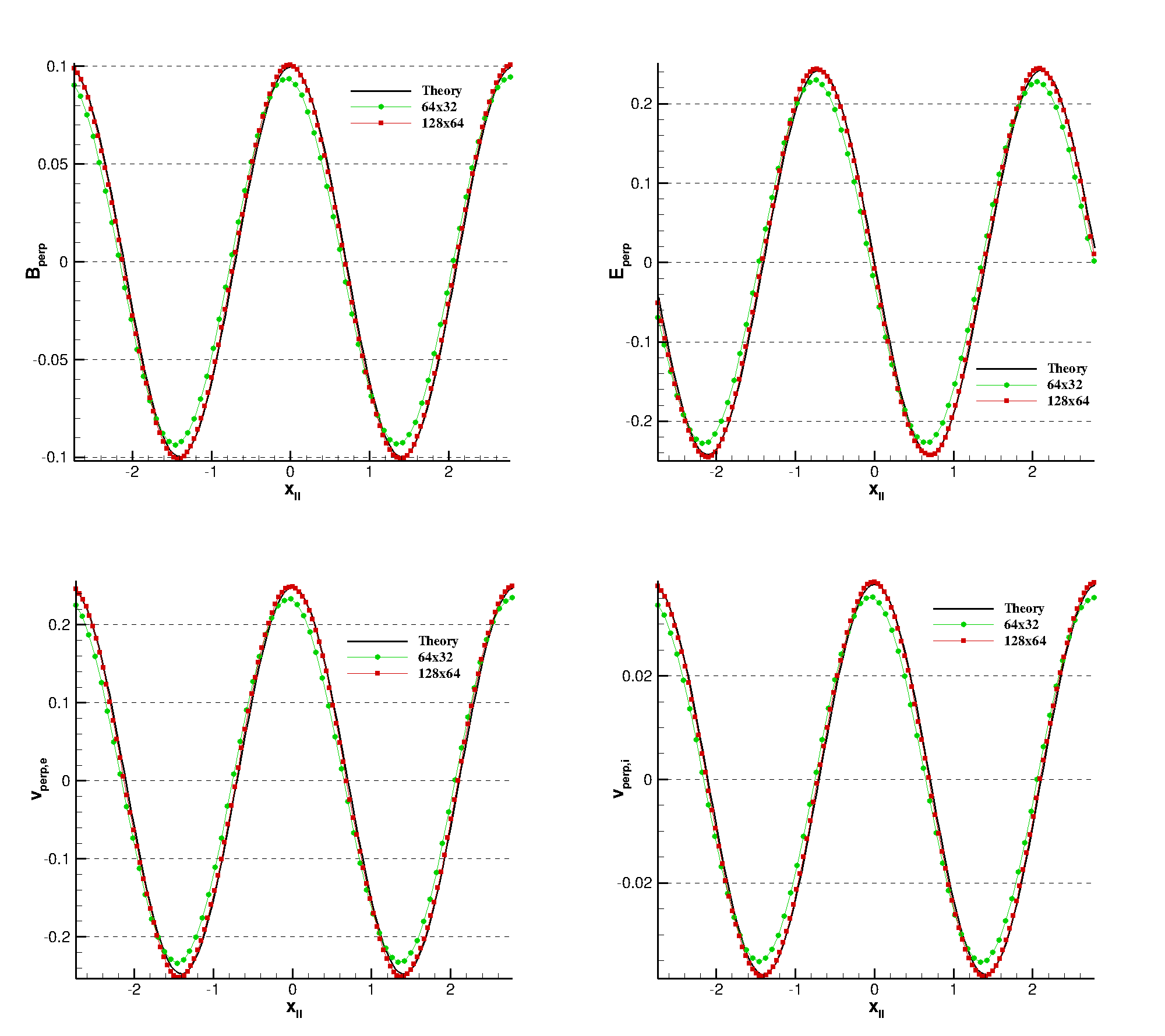}
\caption{\label{fig:CPW} Perpendicular component of the magnetic field (top left), electric field (top right), perpendicular velocity (bottom) of electron (left) and ions (right) for different mesh resolutions ($N=64,128$ indicated as green and red respectively) after five periods. The analytical solution is shown in black. All variables are normalized to their reference values (see \cite{AlvarezLaguna18-CPC}). The solution for $N=256$ is omitted for clarity.}
\end{figure*}

Figure \ref{fig:CPW} shows the perpendicular component of the magnetic and electric fields on the top row (left and right respectively) and the electrons and ions velocities in the bottom row (left and right respectively) at $t=10\pi/\omega$. The solution for both $N=64,128$ are shown as well as the analytical solution (black).

\begin{figure*}[ht!]
\centering
\includegraphics[width=0.5\textwidth]{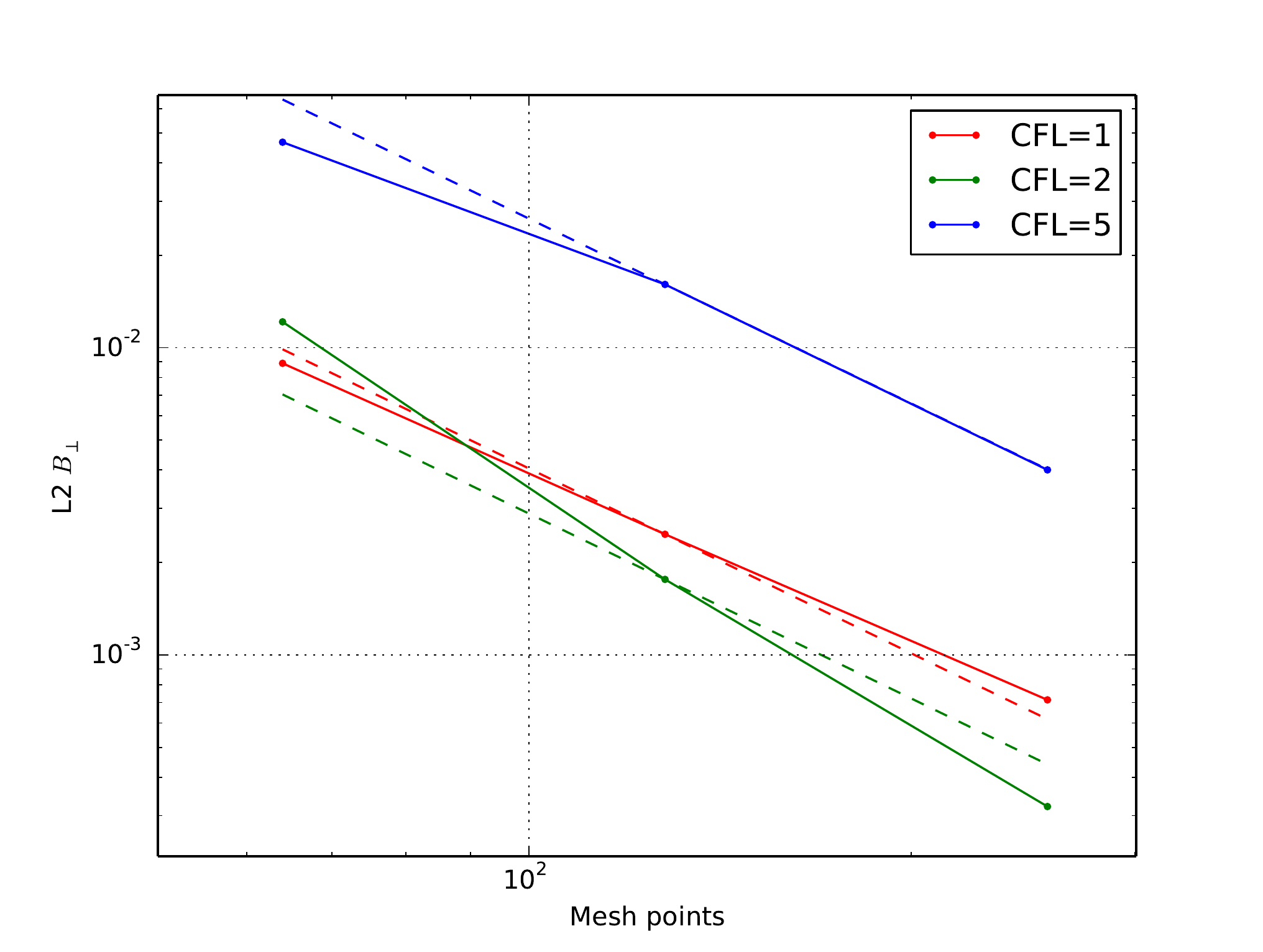}
\caption{ \label{fig:convergenceCFL} L2 error of the perpendicular component of the magnetic field for $N=64,128,256$. The dotted lines represents a second-order error slope.} 
\end{figure*}

 We can also compute the L2 norm to check the convergence of the scheme. Furthermore, as the solver is fully implicit, we can test this convergence for $CFL>1$. In Figure \ref{fig:convergenceCFL} we show the convergence of the perpendicular component of the magnetic field for $CFL=1,2,5$. As expected, we obtain a second order convergence (dashed lines). We can thus confirm that the implementation described above yields correct physical results for this benchmark. Now we can proceed to analyze the performance for these simulations.

\subsubsection{Assembly and solving times}

It is common for implicit codes on the CPU to spend most of the time solving the linear system of equations, this being one of the major drawbacks of those approaches. In this case, however, due to the complexity of the discretized multi-physics model involved, most of the time is spent on the assembly of the system ($70-90 \%$). This is one of the reasons that are considered for porting this part of the code to GPU's.

The first test case is for validation purposes and no major speedup is expected. Small problems indeed do not provide enough instruction to feed the large number of GPU cores. Therefore the GPU runs on sub-optimal regime. The presented benchmarks in the preceding section have, at most, 8192 cells, which may not be enough to fully profit from the GPU parallelization, so we added a finer mesh with $N=256$ to better capture the improvement of the GPU's performance with the number of cells.

For the first metric, the results for the $N=64,128,256$ meshes is shown in Tables \ref{tab:Iter64x32}, \ref{tab:Iter128x64} and \ref{tab:Iter256x128} respectively. For all mesh sizes, when solving with the pure CPU implementation (i.e, PETSC CPU - CPU) most of the time ($70-90 \%$) is spent computing the residual and its jacobian, as expected.

When building the system on the GPU and solving with PETSc on the CPU (i.e. PETSC CPU - GPU configuration) with the Kepler micro-architecture the assembly-to-solving ratio is decreased to $\sim 65/35$ in most cases, whereas for the modern Pascal micro-architecture this ratio is decreased even further, down to $30/70$.

\begin{table*}[h!]
\centering \small
\begin{tabular}{cc||c|c|c|c|c|c|c|c|c|}
\cline{3-10}
\multicolumn{2}{c|}{Mesh:} & \multicolumn{2}{c|}{PETSc CPU - CPU} & \multicolumn{2}{c|}{PETSc CPU - GPU} & \multicolumn{2}{c|}{PETSc GPU - GPU} & \multicolumn{2}{c|}{PARALUTION -  GPU} \\ \cline{3-10}
\multicolumn{2}{c|}{64x32} & Time (s) & Percentage & Time (s) & Percentage & Time (s) & Percentage & Time (s) & Percentage  \\ \hline
\multicolumn{1}{|l|}{\multirow{ 3}{*}{K10}}& Assembly & 0.340 & 90.2\% &  0.090 & 70.3\% & 0.216 & 87.2\% & 0.110 & 28.8\% \\
\multicolumn{1}{|l|}{}                     & Solving  & 0.037 & 9.8\% &  0.038 & 29.7\% & 0.032 & 12.8\% & 0.272 & 71.2\%  \\ \cline{2-10}
\multicolumn{1}{|l|}{}                     & Total    & 0.377 & 100\% &  0.128 & 100\% & 0.248 & 100\% & 0.382 & 100\%    \\ \hline \hline
\multicolumn{1}{|l|}{\multirow{ 3}{*}{K40}}& Assembly & 0.224 & 90.8\% &  0.071 & 67.0\% & 0.141 & 87.8\% & 0.073 & 30.3\% \\
\multicolumn{1}{|l|}{}                     & Solving  & 0.023 & 9.2\%  &  0.035 & 33.0\% & 0.020 & 12.2\% & 0.168 & 69.7\%  \\ \cline{2-10}
\multicolumn{1}{|l|}{}                     & Total    & 0.247 & 100\%  &  0.106 & 100\% &  0.161 & 100\%  & 0.241 & 100\%  \\   \hline \hline
\multicolumn{1}{|l|}{\multirow{ 3}{*}{P100}}& Assembly & 0.369 & 76.7\%  &  0.060 & 34.7\% & 0.097 & 72.9\% & 0.078 & 16.3\% \\
\multicolumn{1}{|l|}{}                      & Solving  & 0.112 & 23.3\%  &  0.112 & 65.3\% & 0.023 & 17.1\% & 0.397 & 83.7\%  \\ \cline{2-10}
\multicolumn{1}{|l|}{}                      & Total    & 0.481  & 100\%  &  0.172 & 100\%  & 0.120 & 100\%  & 0.475 & 100\%   \\   \hline
\end{tabular}
\caption{ Mean times per iteration for assembling the system (i.e. compute pseudo-steady residual and its Jacobian) and solving it for $N=64$.}
\label{tab:Iter64x32}
\end{table*}

If we solve the system on the GPU with PETSc there is no important improvement on the solving time  but for the P100, which performs better than those using the Kepler micro-architecture. However, in general, the time spent building the system is increased due to the different matrix allocation format when solving on the GPU.

\begin{table*}[h!]
\centering \small
\begin{tabular}{cc||c|c|c|c|c|c|c|c|c|}
\cline{3-10}
\multicolumn{2}{c|}{Mesh:} & \multicolumn{2}{c|}{PETSc CPU - CPU} & \multicolumn{2}{c|}{PETSc CPU - GPU} & \multicolumn{2}{c|}{PETSc GPU - GPU} & \multicolumn{2}{c|}{PARALUTION -  GPU} \\ \cline{3-10}
\multicolumn{2}{c|}{128x64} & Time (s) & Percentage & Time (s) & Percentage & Time (s) & Percentage & Time (s) & Percentage  \\ \hline
\multicolumn{1}{|l|}{\multirow{ 3}{*}{K10}}& Assembly & 1.37 & 91.3\% &  0.33 & 70.8\% & 0.67 & 86.9\% & 0.41 & 27.0\% \\
\multicolumn{1}{|l|}{}                     & Solving  & 0.13 & 8.7\% &  0.14 & 29.2\% & 0.10 & 13.1\% & 1.12 & 73.0\%  \\ \cline{2-10}
\multicolumn{1}{|l|}{}                     & Total    & 1.50 & 100\% &  0.47 & 100\% & 0.77 & 100\% & 1.53 & 100\%    \\ \hline \hline
\multicolumn{1}{|l|}{\multirow{ 3}{*}{K40}}& Assembly & 0.91 & 90.6\% &  0.19 & 60.7\% & 0.51 & 87.4\% & 0.19 & 22.8\% \\
\multicolumn{1}{|l|}{}                     & Solving  & 0.10 & 9.4\%  &  0.13 & 39.3\% & 0.07 & 12.6\% & 0.65 & 77.2\%  \\ \cline{2-10}
\multicolumn{1}{|l|}{}                     & Total    & 1.01 & 100\%  &  0.32 & 100\%  & 0.58 & 100\%  & 0.84 & 100\%  \\   \hline \hline
\multicolumn{1}{|l|}{\multirow{ 3}{*}{P100}}& Assembly & 1.48 & 79.0\%  &  0.17 & 28.6\% & 0.31 & 77.6\% & 0.16 & 20.3\% \\
\multicolumn{1}{|l|}{}                      & Solving  & 0.39  & 21.0\% &  0.40 & 71.4\% & 0.09 & 22.4\% & 0.62 & 79.7\%  \\ \cline{2-10}
\multicolumn{1}{|l|}{}                      & Total    & 1.87 & 100\%   &  0.57 & 100\%  & 0.40 & 100\% &  0.78 & 100\%   \\   \hline
\end{tabular}
\caption{ Mean times per iteration for assembling the system (i.e. compute pseudo-steady residual and its Jacobian) and solving it for $N=128$.}
\label{tab:Iter128x64}
\end{table*}

We then solve the system with PARALUTION on the GPU. For the coarsest meshes ($N=64,128$) this increases substantially the solving time, in some cases being longer than with the PETSc CPU - CPU configuration. For $N=256$ this problem is lessened, and PARALUTION solves the system faster than the other configurations. Therefore, for small problem sizes PARALUTION is comparatively slower than PETSc when solving the system on the GPU, but it can give the same performance for medium sized problems ($\sim 30000$ cells).

\begin{table*}[h!]
\centering \small
\begin{tabular}{cc||c|c|c|c|c|c|c|c|c|}
\cline{3-10}
\multicolumn{2}{c|}{Mesh:} & \multicolumn{2}{c|}{PETSc CPU - CPU} & \multicolumn{2}{c|}{PETSc CPU - GPU} & \multicolumn{2}{c|}{PETSc GPU - GPU} & \multicolumn{2}{c|}{PARALUTION -  GPU} \\ \cline{3-10}
\multicolumn{2}{c|}{256x128} & Time (s) & Percentage & Time (s) & Percentage & Time (s) & Percentage & Time (s) & Percentage  \\ \hline
\multicolumn{1}{|l|}{\multirow{ 3}{*}{K10}}& Assembly & 5.61 & 92.5\% &  1.24 & 72.3\% & 2.67 & 87.6\% & 1.46 & 86.4\% \\
\multicolumn{1}{|l|}{}                     & Solving  & 0.46 & 7.5\% &  0.47 & 27.7\% & 0.38 & 12.4\% & 0.23 & 13.6\%  \\ \cline{2-10}
\multicolumn{1}{|l|}{}                     & Total    & 6.07 & 100\% &  1.71 & 100\% & 3.05 & 100\% & 1.69 & 100\%    \\ \hline \hline
\multicolumn{1}{|l|}{\multirow{ 3}{*}{K40}}& Assembly & 3.67 & 91.9\% &  0.72 & 68.6\% & 1.71 & 89.3\% & 0.75 & 86.0\% \\
\multicolumn{1}{|l|}{}                     & Solving  & 0.32 & 8.1\% &  0.33 & 31.4\% & 0.21 & 10.7\% & 0.12 & 14.0\%  \\ \cline{2-10}
\multicolumn{1}{|l|}{}                     & Total    & 3.99 & 100\% &  1.05 & 100\% & 1.92 & 100\% & 0.87 & 100\%    \\ \hline \hline
\multicolumn{1}{|l|}{\multirow{ 3}{*}{P100}}& Assembly & 5.94 & 81.5\% &  0.57 & 29.7\% & 1.19 & 76.4\% & 0.50 & 80.4\% \\
\multicolumn{1}{|l|}{}                      & Solving  & 1.35 & 18.5\% &  1.36 & 70.3\% & 0.37 & 23.6\% & 0.12 & 19.6\%  \\ \cline{2-10}
\multicolumn{1}{|l|}{}                      & Total    & 7.29 & 100\%  &  1.93 & 100\%  & 1.56  & 100\% & 0.62 & 100\%    \\ \hline 
\end{tabular}
\caption{ Mean times per iteration for assembling the system (i.e. compute pseudo-steady residual and its Jacobian) and solving it for $N=256$.}
\label{tab:Iter256x128}
\end{table*}

For these benchmarks, the P100 stands out, and the finest mesh shows more satisfactory results. We have to recall that these timings only take into account the time computing the pseudo-steady residual, its Jacobian and solving the system. In the following we will study the overall speedups, taking into account the whole simulation process.

\subsubsection{Total speedup}

Tables \ref{tab:Speedup64x32}, \ref{tab:Speedup128x64} and \ref{tab:Speedup256x128} show the speedups for $N=64,128,256$ respectively, in a production-like environment. For $N=64$ we can appreciate the effects of low occupancy on the speedups. Although building the system on the GPU improves the performance in all cases (by a factor of $\sim$3x), solving it on the GPU does not further decrease the time-to-solution for the Kepler micro-architecture. Indeed, using the PARALUTION - GPU configuration yields slowdowns for all architectures. Also, we can appreciate the effect discussed before: solving small problems on GPU with PETSc is more efficient than with PARALUTION. Although both LSS wrapper modules are enclosed in general objects (see Sec. \ref{sec:paralution}), each implementation is specific, and this can add appreciable differences in the timings.

\begin{table*}[h!]
\centering
\begin{tabular}{c|c|c|c|c|c|c|c|c|}
\cline{2-9}
Mesh:&\multicolumn{2}{c|}{PETSc CPU - CPU} & \multicolumn{2}{c|}{PETSc CPU - GPU} &  \multicolumn{2}{c|}{PETSc GPU - GPU} & \multicolumn{2}{c|}{PARALUTION -  GPU}\\ \cline{2-9}
64x32 & Time (s) & Speedup & Time (s) & Speedup & Time (s) & Speedup & Time (s) & Speedup\\ \hline
\multicolumn{1}{|c||}{K10} & 5m 42.9s & 1x & 1m 35.7s & 3.58x &   2m 45.1s & 2.08x & 8m 52.0s & 0.64x \\
\multicolumn{1}{|c||}{K40} & 3m 19.4s & 1x  & 1m 12.8s & 2.73x &  1m 50.0s & 1.81x & 13m 43.4s & 0.24x \\
\multicolumn{1}{|c||}{P100}& 6m 27.7s & 1x & 2m 15.1s & 2.87x & 1m 31.2s & 4.25x & 12m 18.1s & 0.52x \\ \hline
\end{tabular}
\caption{Wall time and speedup for $N=64$ with different systems and configurations. In all cases 200 iterations were computed.}
\label{tab:Speedup64x32}
\end{table*}

The relative difference between the solvers, when using the GPU, is decreased for $N=128$. In this case, the P100 shows better results, and the PARALUTION - GPU configuration runs faster than the pure CPU implementation. However, the older Kepler GPU's still are below 1x speedup. This difference between micro-architectures is also noticeable with the PETSc GPU - GPU configuration.

\begin{table*}[h!]
\centering
\begin{tabular}{c|c|c|c|c|c|c|c|c|}
\cline{2-9}
Mesh:&\multicolumn{2}{c|}{PETSc CPU - CPU} & \multicolumn{2}{c|}{PETSc CPU - GPU} &  \multicolumn{2}{c|}{PETSc GPU - GPU} & \multicolumn{2}{c|}{PARALUTION -  GPU}\\ \cline{2-9}
128x64 & Time (s) & Speedup & Time (s) & Speedup & Time (s) & Speedup & Time (s) & Speedup\\ \hline
\multicolumn{1}{|c||}{K10} & 19m 46.6s & 1x & 5m 40.6s & 3.48x &   10m 4.8s & 1.96x & 27m 58.6s & 0.71x \\
\multicolumn{1}{|c||}{K40} & 13m 16.6s & 1x  & 3m 28.6s & 3.81x &   6m 44.0s & 1.97x & 23m 21.0s & 0.56x \\
\multicolumn{1}{|c||}{P100}& 24m 37.7s & 1x &  6m 56.4s & 3.52x & 5m 14.5s & 4.67x & 16m 4.4s & 1.52x \\ \hline
\end{tabular}
\caption{Wall time and speedup for $N=124$ with different systems and configurations.}
\label{tab:Speedup128x64}
\end{table*}

When studying the finest mesh ($N=256$), the GPU's occupancies are high enough to improve the performance substantially for all micro-architectures. PARALUTION starts to outperform PETSc and the maximum speedup is given by the P100, running up to $9.5$ times faster.

\begin{table*}[h!]
\centering
\begin{tabular}{c|c|c|c|c|c|c|c|c|}
\cline{2-9}
Mesh:&\multicolumn{2}{c|}{PETSc CPU - CPU} & \multicolumn{2}{c|}{PETSc CPU - GPU} &  \multicolumn{2}{c|}{PETSc GPU - GPU} & \multicolumn{2}{c|}{PARALUTION -  GPU}\\ \cline{2-9}
256x128 & Time (s) & Speedup & Time (s) & Speedup & Time (s) & Speedup & Time (s) & Speedup\\ \hline
\multicolumn{1}{|c||}{K10} & 1h 19m 28.3s & 1x & 20m 3.3s & 3.96x &   38m 47.0s & 2.05x & 21m 3.0s & 3.77x \\
\multicolumn{1}{|c||}{K40} & 1h 2m 34.1s & 1x & 14m 13.9s & 4.39x &   26m 41.4s & 2.34x & 24m 17.6s & 2.54x \\
\multicolumn{1}{|c||}{P100} & 1h 33m 55.4s & 1x & 22m 43.6s & 4.13x &   20m 26.8s & 4.59x & 9m 43.7s & 9.65x \\ \hline
\end{tabular}
\caption{Wall time and speedup for $N=256$ with different systems and configurations.}
\label{tab:Speedup256x128}
\end{table*}

In all meshes studied before, just computing the residual and its Jacobian on the GPU gives a steady $3\sim4$ times speedups for all micro-architectures. Although it is a favorable result, it also indicates that more work can be done to take advantage of newer micro-architectures. In the following we will study another case to explore finer meshes and complex simulations.


\subsection{GEM challenge}
In order to assess if the GPU-enabled implicit solver that has been described above obtains the same numerical results than the previous CPU implementation even for more demanding simulations, we simulate the Geospace Environment Modeling (GEM) challenge \cite{Birn01-GEM}. We then compare our results with those obtained by Alvarez Laguna et al.~\cite{AlvarezLaguna18-CPC}. This case aims at studying the magnetic reconnection that takes place in the magnetotail with the ion-electron model described in this work.

The initial configuration is formed by a Harris equilibrium with a perturbation on the magnetic field in order to start the reconnection from the non-linear regime. Detailed descriptions of the initial conditions can be found in ~\cite{AlvarezLaguna18-CPC,Birn01-GEM}. In this simulation, we set the mass ratio $m_i/m_e = 25$, the temperature ratio at the initial step $T_e/T_i = 0.2$, and for both fluids an ideal gas law is assumed with $\gamma=5/3$.

\subsubsection{Numerical results}

The physical domain consists of a square, where a double current sheet is simulated. By doing this, we can define periodic boundary conditions in all sides. The mesh is formed by 512x512 elements in a Cartesian grid, however, also a coarser mesh of 256x256 elements was used for benchmarking. The same time step as in \cite{AlvarezLaguna18-CPC} is chosen.

For the comparison with previous results, a simulation up to $t \Omega_{ci} =25$ (i.e. 10000 iterations) was carried out using a single P100 GPU (and its corresponding CPU core) using the newly implemented PARALUTION GPU configuration. Although it will be discussed in details below, it is worth mentioning that this simulation took 49h 16m. 

\begin{figure*}[!ht]
	\centering
    \begin{subfigure}{.5\textwidth}
    \includegraphics[width=1.\linewidth]{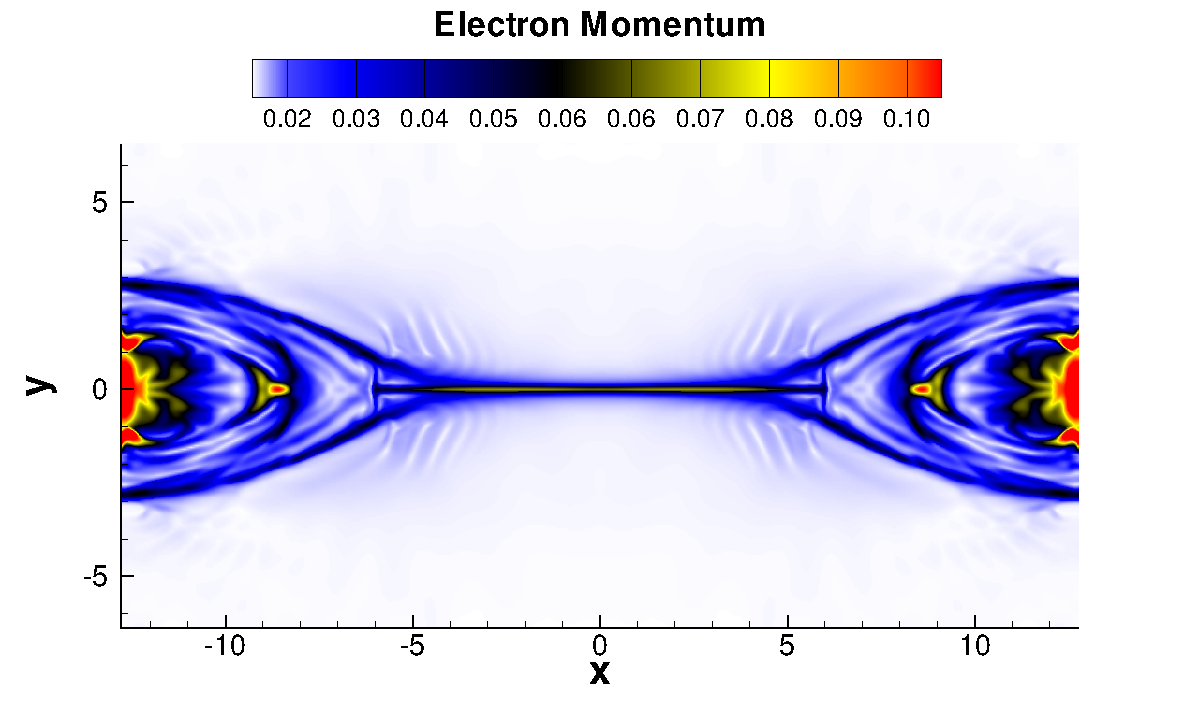}
    \end{subfigure}%
	\begin{subfigure}{.5\textwidth}
    \includegraphics[width=1.\linewidth]{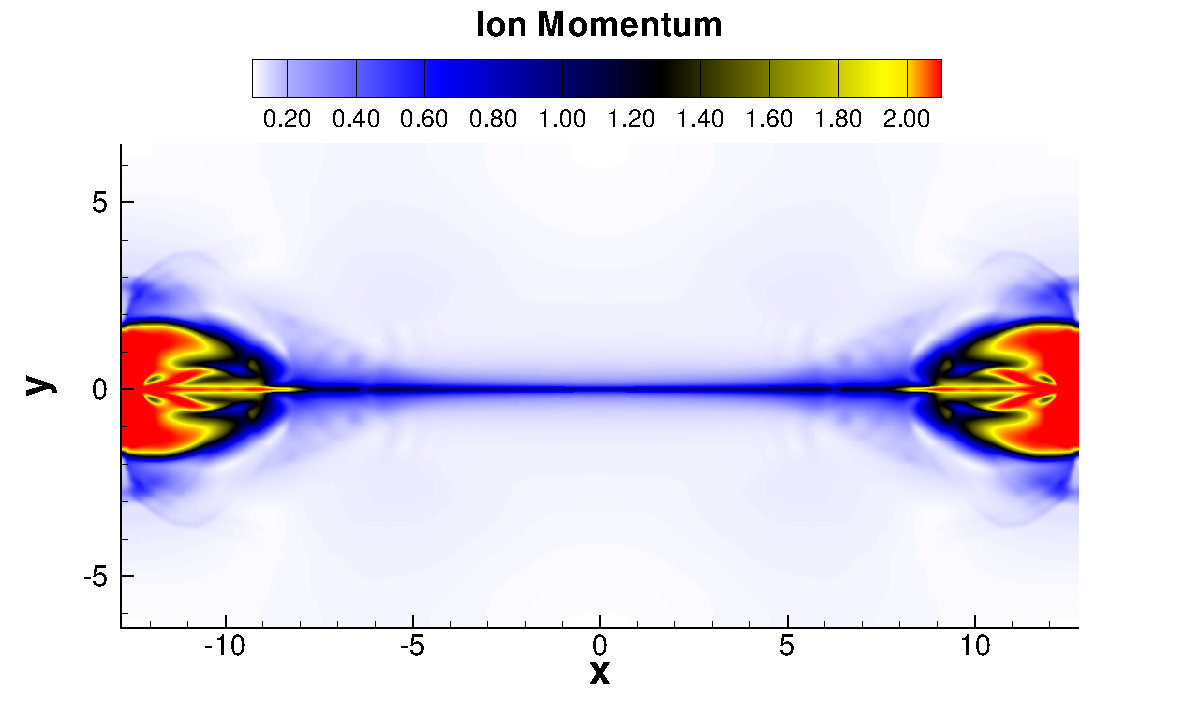}
	\end{subfigure}
    
	\begin{subfigure}{.5\textwidth}
	\centering
    \includegraphics[width=1.\textwidth]{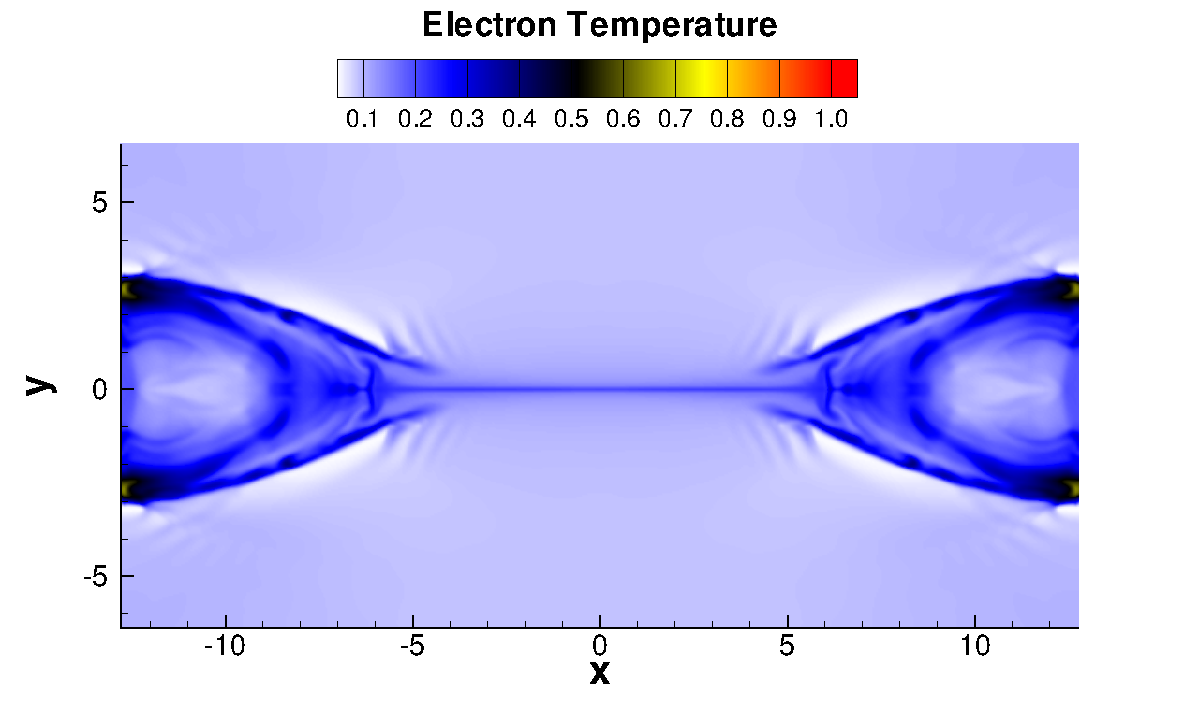}
	\end{subfigure}%
	\begin{subfigure}{.5\textwidth}
	\centering
    \includegraphics[width=1.\textwidth]{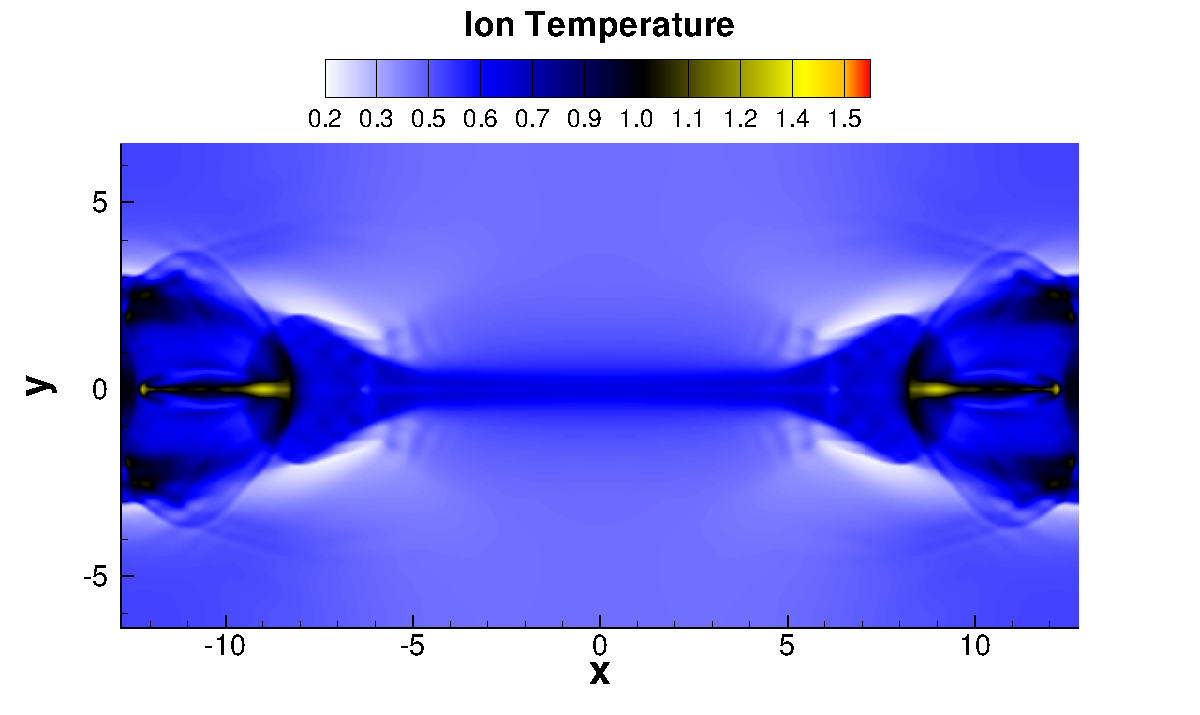}
    \end{subfigure}

    \caption{\label{fig:comparative_results} Momentum (top row) of ions (right) and electrons (left) and their temperatures (bottom row) at $t \Omega_{ci} =25$. Both variables are normalized to the ions initial values. This figure is to be compared with Figure 10 of \cite{AlvarezLaguna18-CPC}. }
\end{figure*}

\begin{figure*}[ht!]
\centering
\includegraphics[width=0.5\textwidth]{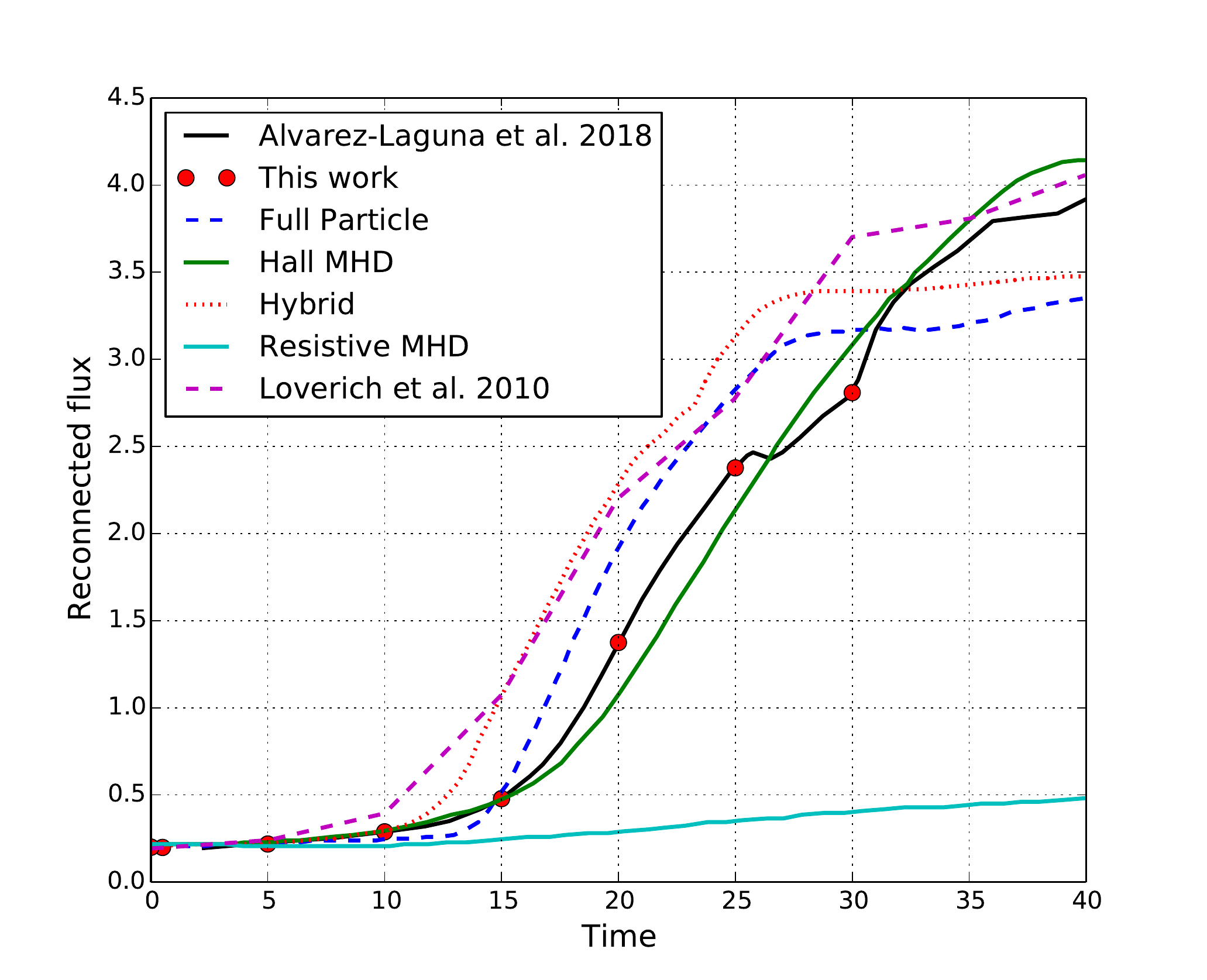}
\caption{ \label{fig:reconnectedflux} Reconnected flux as a function of time (in ion cyclotron periods). The same model (described in \cite{AlvarezLaguna18-CPC}) is shown in black. As expected, the new GPU-enabled solver does not change this result. Furthermore results reported by \cite{Birn01-GEM} and \cite{loverich11} are shown for comparison.} 
\end{figure*}

Figure \ref{fig:comparative_results} shows the momentum (top row) and temperature (bottom) from both electrons and ions. The results show the same flow field than this of \cite{AlvarezLaguna18-CPC} in all scales,  thus qualitatively the GPU-enabled solver yields the same results. We can also measure the reconnected flux as prescribed in \cite{Birn01-GEM} to quantitatively assess the correctness of the solution. This is shown in Figure \ref{fig:reconnectedflux} and as expected we recover the same solution as the CPU-only version of this solver.

Now that we have verified the implementation, we can proceed to analyze the obtained speedups for this simulation. Before continuing with the timings, we shall notice that due to the limited memory of the K10 GPU, the finest mesh was not used in that system. Furthermore, due to limits on the allocation of the Jacobian matrix, the configurations PETSC CPU/GPU - GPU were not tested with the fine mesh.

\subsubsection{Assembly and solving times}

The results for the first metric are shown in Tables \ref{tab:Iter256x256} and \ref{tab:Iter512x512} for the coarse (256x256 cells) and the fine (512x512 cells) mesh. In the same fashion than the preceding benchmarks, using the pure CPU implementation, most of the time is spent assembling the system. If this is offloaded to the GPU, the ratio of assembly to solving times decreases to $\sim 65/35$ for the Kepler GPU's and to $30/70$ using the P100.

\begin{table*}[h!]
\centering \small
\begin{tabular}{cc||c|c|c|c|c|c|c|c|c|}
\cline{3-10}
\multicolumn{2}{c|}{Mesh:} & \multicolumn{2}{c|}{PETSc CPU - CPU} & \multicolumn{2}{c|}{PETSc CPU - GPU} & \multicolumn{2}{c|}{PETSc GPU - GPU} & \multicolumn{2}{c|}{PARALUTION -  GPU} \\ \cline{3-10}
\multicolumn{2}{c|}{256x256} & Time (s) & Percentage & Time (s) & Percentage & Time (s) & Percentage & Time (s) & Percentage  \\ \hline
\multicolumn{1}{|l|}{\multirow{ 3}{*}{K10}}& Assembly & 14.78 & 92.8\% &  2.65 & 69.8\% & 5.33 & 87.1\% & 2.77 & 86.4\% \\
\multicolumn{1}{|l|}{}                     & Solving  & 1.14 & 7.2\% &  1.14 & 30.2\% & 0.79 & 12.9\% & 0.43 & 13.6\%  \\ \cline{2-10}
\multicolumn{1}{|l|}{}                     & Total    & 15.92 & 100\% &  3.79 & 100\% & 6.12 & 100\% & 3.20 & 100\%    \\ \hline \hline
\multicolumn{1}{|l|}{\multirow{ 3}{*}{K40}}& Assembly & 8.87 & 89.3\% &  1.52 & 62.1\% & 3.47 & 89.0\% & 1.51 & 86.2\% \\
\multicolumn{1}{|l|}{}                     & Solving  & 1.07 & 10.7\%  &  0.93 & 37.9\% & 0.43 & 11.0\% & 0.24 & 13.8\%  \\ \cline{2-10}
\multicolumn{1}{|l|}{}                     & Total    & 9.94 & 100\%&  2.45 & 100\% & 3.90 & 100\% & 1.75 & 100\%  \\   \hline \hline
\multicolumn{1}{|l|}{\multirow{ 3}{*}{P100}}& Assembly & 11.89 & 81.4\% &  1.05 & 27.8\% & 2.30 & 75.8\% & 0.92 & 83.1\% \\
\multicolumn{1}{|l|}{}                      & Solving  & 2.72  & 18.6\% &  2.73 & 72.2\% & 0.74 & 24.2\% & 0.19 & 16.9\%  \\ \cline{2-10}
\multicolumn{1}{|l|}{}                      & Total   & 14.61 & 100\%   &  3.78 & 100\%  & 3.04 & 100\% & 1.11 & 100\%   \\   \hline
\end{tabular}

\caption{ Mean times taken per iteration to assemble the system (i.e. compute pseudo-steady residual and its Jacobian) and solve it, for the coarse mesh (i.e 256x256).}
\label{tab:Iter256x256}
\end{table*}

The ratios when solving the system on the GPU with PETSc are consistent with those obtained before. Compared with the previous benchmarks, we can now clearly see the advantage of using PARALUTION over PETSc, now that we have enough elements. The final assembly-to-solve ratio is close to $\sim 85/15$, similar to that from the PETSc CPU - CPU configuration. The problem is still dominated by the computation of the residual and its Jacobian, thus there is still room for further optimization in this computations. Nevertheless it is important to notice that, although they have similar ratios, the total time taken is substantially smaller with the PARALUTION - GPU configuration.

\begin{table*}[h!]
\centering \small
\begin{tabular}{cc||c|c|c|c|}
\cline{3-6}
\multicolumn{2}{c|}{Mesh:}& \multicolumn{2}{c|}{PETSc CPU - CPU} & \multicolumn{2}{c|}{PARALUTION -  GPU} \\ \cline{3-6}
\multicolumn{2}{c|}{512x512} & Time (s) & Percentage & Time (s) & Percentage \\  \hline
\multicolumn{1}{|l|}{\multirow{ 3}{*}{K40}}& Assembly & 30.00 & 90.4\%  & 5.37 & 84.9\% \\
\multicolumn{1}{|l|}{}                     & Solving  & 3.18 & 9.6\%    & 0.96 & 15.1\%  \\ \cline{2-6}
\multicolumn{1}{|l|}{}                     & Total    & 33.18 & 100\%   & 6.33 & 100\%        \\   \hline \hline
\multicolumn{1}{|l|}{\multirow{ 3}{*}{P100}}& Assembly & 47.49 & 76.8\%    & 3.50 & 87.5\% \\
\multicolumn{1}{|l|}{}                      & Solving  & 14.32 & 23.2\%    & 0.50 & 12.5\%  \\ \cline{2-6}
\multicolumn{1}{|l|}{}                      & Total    & 61.81 & 100\%      & 4.00 & 100\%   \\   \hline
\end{tabular}
\caption{ Mean times taken per iteration to assemble the system (i.e. compute pseudo-steady residual and its Jacobian) and solve it, for the fine mesh (i.e 512x512).}
\label{tab:Iter512x512}
\end{table*}

\subsubsection{Total speedup} 

Table \ref{tab:Speedup256x256} shows the total wall time for the three different systems, using the four configurations and the coarse mesh. As briefly described before, there are noticeable differences in the GPU's micro-architectures. Both using Kepler yield similar speedups ($\sim$4x), whereas the P100 yields up to 12x. This difference is directly associated with the recent progress in GPU's technology. The K10 GPU was launched in 2013, and the P100 in 2016, so there has been a huge improvement in the memory bandwidth of the GPU's, their clock speed and number of CUDA cores (see Table \ref{tab:GPUs}).

\begin{table*}[h!]
\centering
\begin{tabular}{c|c|c|c|c|c|c|c|c|}
\cline{2-9}
Mesh &\multicolumn{2}{c|}{PETSc CPU - CPU} & \multicolumn{2}{c|}{PETSc CPU - GPU} &  \multicolumn{2}{c|}{PETSc GPU - GPU} & \multicolumn{2}{c|}{PARALUTION -  GPU}\\ \cline{2-9}
256x256 & Time (s) & Speedup & Time (s) & Speedup & Time (s) & Speedup & Time (s) & Speedup\\ \hline
\multicolumn{1}{|c||}{K10} & 2h 41m 17s & 1x & 42m 5s & 3.83x &  1h 19m 6s & 2.04x & 38m 45s & 4.16x \\
\multicolumn{1}{|c||}{K40} & 1h 58m 2s & 1x  & 27m 22s & 4.31x &   51m 13s & 2.30x & 27m 3s & 4.36x \\
\multicolumn{1}{|c||}{P100}& 3 h 7m 19s & 1x & 45m 23s & 4.13x & 29m 56s & 6.26x & 15m 10s & 12.35x \\ \hline
\end{tabular}
\caption{Wall time and speedup for the coarse mesh in different systems. In all cases 200 iterations were computed.}
\label{tab:Speedup256x256}
\end{table*}

\begin{table*}[h!]
\centering
\begin{tabular}{c|c|c|c|c|}
\cline{2-5}
Mesh: &\multicolumn{2}{c|}{PETSc CPU - CPU} & \multicolumn{2}{c|}{PARALUTION -  GPU}\\ \cline{2-5}
512x512 & Time (s) & Speedup & Time (s) & Speedup \\ \hline
\multicolumn{1}{|c||}{K40} & 8h 1m 6s & 1x & 1h 50m 26s & 4.36x  \\
\multicolumn{1}{|c||}{P100}  & 13h 59m 54s & 1x  &  59m 1s & 14.23x \\ \hline
\end{tabular}
\caption{ Wall time and speedup for the fine mesh in different systems.}
\label{tab:Speedup512x512}
\end{table*}

For the fine mesh (Table \ref{tab:Speedup512x512}) the obtained speedup improves for both micro-architectures, and again the P100 stands out with a remarkable 14.2x speedup. However, the PETSc CPU - CPU configuration appears to be less efficient on the P100 system\footnote{This could be explained by the lower frequency of the CPU and its smaller cache size.}. In any case, we can naively compare the best time-to-solution using the K40 (i.e 8h 1m 16s) and the time taken on the P100 using the PARALUTION - GPU configuration, yielding a theoretical speedup of 8.15x. This result is still remarkable for an implicit solver.

The speedup for the K40 shows no further improvement when increasing the mesh size, indicating that the GPU has reached the maximum achievable performance with the current implementation. This is however not the case for the P100, so a slightly finer mesh could further increase the speedup. 

\begin{figure}[ht!]
\centering
\includegraphics[width=0.5\textwidth]{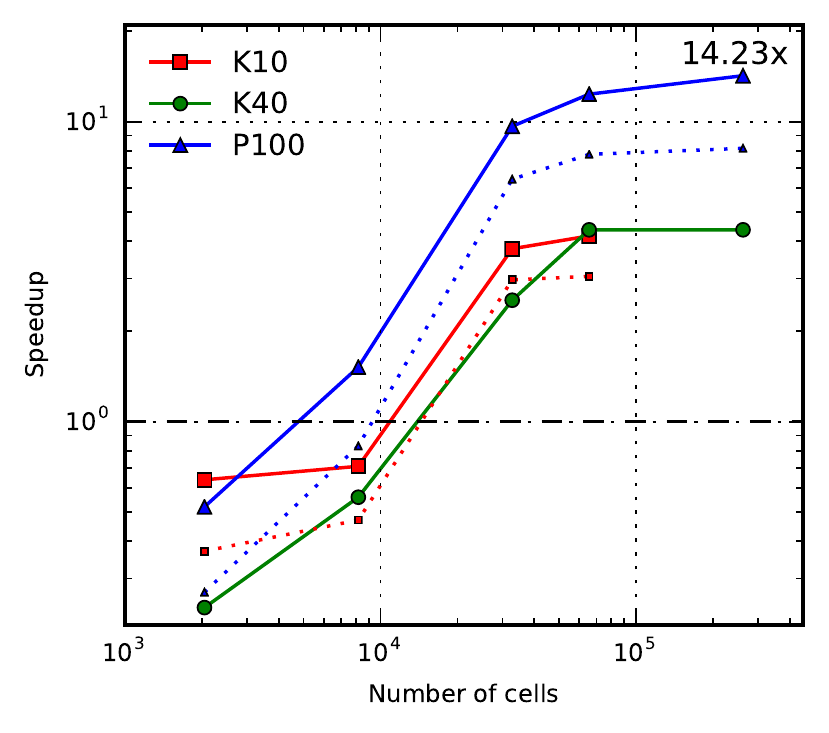}
\caption{\label{fig:summary} Speedups for the PARALUTION - GPU configuration on the three different systems as a function of the number of cells. The dotted lines indicate the theoretical speedup comparing with the K40 system, which in general yielded the best PETSc CPU - CPU timings. The horizontal dashed lines indicate where the speedup equals 1x, i.e, the wall time is the same for both configurations.}
\end{figure}

In Figure \ref{fig:summary} we show the speedups for the PARALUTION - GPU configuration for all the different meshes tested before, in order to better visualize the previous discussion. Although the studied physical cases are different, the overall behavior of the scaling is coherent and thus can be shown together.

We can clearly see different phases of the GPU performance as defined by \cite[p.~70]{AissaThesis17}. For low number of cells ($\lesssim$ 15000 cells), a exponential growth of the speedup is observed, which then tends to saturate. This peak speedup depends on the GPU's microarchitecture, and so does the number of cells to reach it. For the older K10 GPU, this saturation point starts with smaller problems, whereas the P100 is barely saturated for the biggest mesh.

Also, the theoretical comparison with the PETSc CPU - CPU timings on the K40 is shown (dotted lines) indicating that even with this more demanding comparison the P100 speedups are always higher than those on the K40. Thus we can expect, in general, up to two times better performance using the P100 GPU instead of the K40 for the current implementation.

\section{Conclusion}

A new implementation for a GPU-enabled multi-fluid solver within COOLFluiD has been described and tested with unsteady cases, showing a significant improvement in the performance. Starting from the basic equations and the numerical methods, the core of the work (residual and Jacobian computation on the GPU, plus full assembly on the GPU with PARALUTION) has been described. This solver is, to the Author's knowledge, the first fully implicit multi-fluid solver ported to GPU's so far.

Extensive tests have been done to study the achieved speedup with different mesh sizes and Linear System Solvers (PETSc and PARALUTION). We have encountered that for small problem sizes ($\lesssim$ 15000 cells)) solving the system on the CPU using PETSc yields the best time-to-solution. On the other hand, for fine meshes ($\gtrsim 30000$ cells), building the system on the GPU and solving it with PARALUTION is faster than with PETSc. We have also tested three different NVIDIA GPU's (K10, K40 and P100), and found that the modern Pascal architecture yields better performance, as expected. In all cases presented here, the simple fact of computing the residual and its Jacobian on the GPU yields a performance boost of 3-4x.

The code has still some limitations, mainly due to the small available memory at GPU's as compared to RAM's, but we expect that in the next years those limitations will be overcome by the constant technology improvement. The latter is already visible in this work, where GPU's launched with 3-years difference show varied results, and the most modern of those can achieve up to a 14.2x speedup.

We think that GPU's are of great value for hydrodynamical simulations, even for implicit and unstructured codes as the one studied in this work, thus we expect more works in different fields of Physics and Computational Science aimed to take full advantage of modern GPU's. It is also of special interest to further develop GPU-enabled linear system solvers, e.g. PARALUTION, as they can be the bottleneck for implicit solvers.

Further work can be carried out in order to extend this implementation to the diffusive fluxes (in order to fully port the COOLFluiD plasma-neutral model in \cite{AlvarezLaguna16}). Nevertheless, this requires a considerable re-structuring of the FVM-CUDA module, and thus was not developed for this work. Moreover, further optimization and fine-tuning could be possible for the residual and Jacobian computation. Next generation NVIDIA GPU's (using the Volta micro-architecture) could be tested with the discussed implementations to characterize the obtained speedups.

\section{Acknowledgments}

The author acknowledges helpful discussions with Mag Selwa. For the P100 system, the authors used the computational resources and services provided by the VSC (Flemish Supercomputer Center), funded by the Research Foundation - Flanders (FWO) and the Flemish Government – department EWI. These results were obtained in the framework of the projects
GOA/2015-014 (KU Leuven), G.0A23.16N (FWO-Vlaanderen) and C$\sim$90347 (ESA Prodex).

\bibliographystyle{elsarticle-num}
\bibliographystyle{plain}
\bibliography{referencesIAA}

\end{document}